\documentclass[pdflatex,sn-mathphys-num]{sn-jnl}% Math and Physical Sciences Numbered Reference Style
\usepackage{graphicx}%
\usepackage{multirow}%
\usepackage{amsmath,amssymb,amsfonts}%
\usepackage{amsthm}%
\usepackage{mathrsfs}
\usepackage[title]{appendix}%
\usepackage{xcolor}%
\usepackage{textcomp}%
\usepackage{manyfoot}%
\usepackage{booktabs}%
\usepackage{algorithm}%
\usepackage{algorithmicx}%
\usepackage{algpseudocode}%
\usepackage{listings}%
\usepackage{abstract}
\usepackage{soul}
\usepackage{xcolor}
\soulregister\cite7
\soulregister\ref7

\theoremstyle{thmstyleone}%
\theoremstyle{thmstyletwo}%
\theoremstyle{thmstylethree}%
\begin{document}

\title[Article Title]{Stabilization of Tetragonal Phase and Aluminum-Doping Effect in a Bilayer Nickelate}


\author*[1]{Jia-Yi Lu}\email{lujiayi0522@zju.edu.cn}
\equalcont{These authors contributed equally to this work.}

\author[1]{Yi-Qiang Lin}%\email{yiqianglin@zju.edu.cn}
\equalcont{These authors contributed equally to this work.}

\author[2]{\fnm{Kai-Xin} \sur{Ye}}%\email{kaixin$\_$ye@zju.edu.cn}
\equalcont{These authors contributed equally to this work.}

\author[1]{Xin-Yu Zhao}

\author[1]{Jia-Xin Li}

\author[2]{\fnm{{Ya-Nan}} \sur{Zhang}}

\author[3]{\fnm{Hao} \sur{Li}}

\author[3]{Bai-Jiang Lv}

\author*[2,4,5]{\fnm{Hui-Qiu} \sur{Yuan}}\email{hqyuan@zju.edu.cn}

\author*[1,5,6]{Guang-Han Cao}\email{ghcao@zju.edu.cn}

\affil[1]{School of Physics, Zhejiang University, Hangzhou 310058, China}

\affil[2]{Center for Correlated Matter and School of Physics, Zhejiang University, Hangzhou 310058, China}

\affil[3]{Key Laboratory of Neutron Physics and Institute of Nuclear Physics and Chemistry, China Academy of Engineering Physics, Mianyang, 621999, China}

\affil[4]{Institute for Advanced Study in Physics, Zhejiang University, Hangzhou 310058, China}

\affil[5]{Institute of Fundamental and Transdisciplinary Research, and State Key Laboratory of Silicon and Advanced Semiconductor Materials, Zhejiang University, Hangzhou 310058, China}

\affil[6]{Collaborative Innovation Centre of Advanced Microstructures, Nanjing University, Nanjing, 210093, China}

\maketitle

\begin{abstract}
	
Recent studies suggest that the tetragonal phase of the Ruddlesden-Popper (RP) bilayer nickelate, La$_3$Ni$_2$O$_7$ or La$_2$PrNi$_2$O$_7$, which is stabilized under high pressures, is responsible for high-temperature superconductivity (HTSC). In this context, realization of the tetragonal phase at ambient pressure could be a rational step to achieve the goal of ambient-pressure HTSC in the nickelate system. By employing the concept of Goldschmidt tolerance factor, we succeed in stabilizing the tetragonal phase by aluminum doping together with post annealing under moderately high oxygen pressure. X-ray and neutron diffractions verify the tetragonal $I4/mmm$ structure for the post-annealed samples La$_3$Ni$_{2-x}$Al$_x$O$_{7-\delta}$ (0.3 $\leq x \leq$ 0.5). The Al-doped samples, including the tetragonal ones, show semiconducting properties, carry localized magnetic moments, and exhibit spin-glass-like behaviors at low temperatures, all of which can be explained in terms of charge carrier localization. Furthermore, high-pressure resistance measurements on post-annealed samples reveal that even a low Al doping ($x$ = 0.05) suppresses superconductivity almost completely. This work gives information about the effect of nonmagnetic impurity on metallicity as well as superconductivity in bilayer nickelates, which would contribute to understanding the superconducting mechanism in RP nickelates.
\end{abstract}

\section{Introduction}

The recent discovery of high-$T_\mathrm{c}$ superconductivity (HTSC) at about 80 K in the pressurized bilayer perovskite nickelate, La$_3$Ni$_2$O$_7$, represents a remarkable advance in the field of superconductivity research~\cite{Sun2023,2024PRX.ChengJG,2024NP.YuanHQ}. Unlike the infinite-layer nickelate superconductor discovered in 2019~\cite{Li2019}, where the electronic configuration of nickel is nearly 3$d^9$ (Ni$^{1+}$), the bilayer nickelate has an electronic configuration of 3$d^{7.5}$ (Ni$^{2.5+}$). The apical oxygen site between the NiO$_2$ bilayers is almost fully occupied, in contrast with the absence of apical oxygen in the infinite-layer system. The oxygen occupancy not only greatly alters the Ni valence state, but also gives rise to strong interlayer coupling within the bilayers, the latter of which is widely accepted to be crucial for the emergence of HTSC~\cite{Sun2023,2023CPL.ZhangGM,Lechermann2023,2023PRB.Dagotto,2023PRB.YangYF,YangF2023.PRL,2024PRL.Kuroki,Geisler2024,2024npj.Geisler-2,2024WuCJ.PRL,SuG2024.PRL}. Currently, one of the major challenges in the research of nickelate superconductors is to realize bulk HTSC at ambient pressure, such that in-depth experimental investigations on the HTSC can be carried on. Note that while the pressure-quench protocol has successfully stabilized the high-pressure superconducting phase in Bi$_{0.5}$Sb$_{1.5}$Te$_3$~\cite{2024PNAS.Deng}, this approach has not yet been replicated in nickelate superconductors~\cite{2024AEM.Xu}.

The bilayer nickelate La$_3$Ni$_2$O$_7$ is an $n=2$ member of the Ruddlesden-Popper (RP) series~\cite{Drennan1982,Zhang1994,SREEDHAR1994,Taniguchi1995,Ling2000}, La$_{n+1}$Ni$_n$O$_{3n+1}$, where $n$ is the number of perovskite LaNiO$_3$ layers in between rocksalt-type LaO layers. At ambient pressure, the NiO$_6$ octahedra in La$_3$Ni$_2$O$_7$ are distorted and rotated, resulting in an orthorhombic structure with space group $Amam$ (No. 63) (this space group is based on a nonstandard setting, such that the $c$ axis retains to be the longest one, consistent with the longest $c$ axis in the tetragonal structure)~\cite{Ling2000}. Upon applying pressure, La$_3$Ni$_2$O$_7$ undergoes phase transitions to another orthorhombic structure ($Fmmm$, No. 69)~\cite{Sun2023} and further to a tetragonal structure (space group $I4/mmm$, No. 139)~\cite{Structure2024,327.100GPa}. Importantly, the out-of-plane Ni$-$O$-$Ni bond angle of those high-pressure stabilized structures turns out to be 180$^{\circ}$, which effectively enhances the interlayer coupling. Moreover, a structural transition from $Amam$ directly to $I4/mmm$ at about 11 GPa was observed in La$_2$PrNi$_2$O$_7$~\cite{Wang2024}, concurrently with the emergence of superconductivity. A similar structural transformation into $I4/mmm$ tetragonal phase accompanying with appearance of bulk superconductivity is also observed in the trilayer nickelate La$_4$Ni$_3$O$_{10}$~\cite{Zhu2024}, which further corroborates the close relationship between crystal structure and HTSC. Very recently, it was found that compressively strained bilayer-nickelate thin films, which show superconductivity at ambient pressure~\cite{APSC.Hwang-1,Zhou2025.nature,APSC.Hwang-2}, also adopt a tetragonal structure~\cite{Zhou2025.nature}. All the information above stimulates the research towards stabilization of the tetragonal structure, which at least serves as an initial step to ultimately realize ambient-pressure HTSC in $\emph{bulk samples}$ of the bilayer nickelate.

The concept of Goldschmidt tolerance factor~\cite{Goldschmidt1926}, defined by $\tau = \frac{r_\mathrm{La} +r_\mathrm{O}}{\sqrt{2}(r_\mathrm{Ni}+r_\mathrm{O})}$, where $\tau$ refers to the ionic radii of the constituent ions, gives a clue to obtain the target tetragonal phase. In the view of crystal chemistry, the tolerance factor basically describes the stability as well as the tendency of lattice distortion in the perovskite-like block layers~\cite{2016CM.tolerance}. Using the effective ionic radii~\cite{Shannon1976}, and taking $r_\mathrm{Ni^{2.5+}}=(r_\mathrm{Ni^{2+}}+r_\mathrm{Ni^{3+}})/2=0.625$ \AA~ for La$_3$Ni$_2$O$_7$, one obtains $\tau=$ 0.916, suggesting that the perovskite-like layers are significantly compressed by the rocksalt LaO layers, which accounts for the orthorhombic distortion at ambient pressure. If the $\tau$ value increases towards 1.0 through suitable chemical substitutions, one expects that the tetragonal phase could be stabilized at ambient pressure. Indeed, a tetragonal bilayer RP phase in Sr-Ni-O system was realized through Al doping~\cite{Inga2018,Yilmaz2024}, and via high-pressure synthesis~\cite{yamane2024}, where the $\tau$ value is nearly 1.0. The theoretical calculations~\cite{Rhodes2024} suggest that the $Fmmm$ structure can be stabilized at ambient pressure by replacing the La$^{3+}$ ion with a larger cation, either Ba$^{2+}$ or Ac$^{3+}$. Ac$_3$Ni$_2$O$_7$ is also predicted to crystallize in the tetragonal $I4/mmm$ structure~\cite{wu2024}. Unfortunately, the element Ac is almost unavailable because of its strong radioactivity and, on the other hand, the allowed solubility limit of Ba in La$_3$Ni$_2$O$_7$ is too small to increase the $\tau$ value appreciably~\cite{Zhang1994-2,2024Bozovic}. Considering the efficiency of Al doping in stabilizing La-Ni-O trilayer structures~\cite{periyasamy2021} and its substantial doping capacity in related RP phases~\cite{Inga2018,periyasamy2021}, partial substitution of Ni sites with smaller cations Al$^{3+}$ ($r_\mathrm{Al^{3+}}=0.535$ \AA) might be a viable approach. Although the Al-doping definitely induces disorder, it still can serve as a probe to investigate the effect of nonmagnetic disorder on Anderson localization as well as potential superconductivity.

In this paper, we report our successful realization of the tetragonal phase in the bilayer nickelate La$_3$Ni$_{2-x}$Al$_x$O$_{7-\delta}$. Single-phase samples were obtained in the Al doping range of $0\leq x\leq 0.5$. While all the as-prepared samples remain orthorhombic, the orthorhombicity tends to decrease with the Al doping. After post annealing under moderately high oxygen pressure, strikingly, the samples with $0.3\leq x\leq 0.5$ finally become tetragonal. The detailed crystal structure of the typical samples of $x=0.4$ was determined by X-ray and neutron powder diffractions. Transport and magnetic measurements under ambient pressure demonstrate that Al doping induces strong carrier localization. High-pressure resistance studies reveal a rapid suppression of superconductivity in La$_3$Ni$_{2-x}$Al$_x$O$_{7-\delta}$ ($x$ = 0.05, 0.1, and 0.4).

\section{RESULTS AND DISCUSSION}\label{sec2}

Polycrystalline samples of La$_3$Ni$_{2-x}$Al$_x$O$_{7-\delta}$ ($x$ = 0, 0.05, 0.1, 0.2, 0.3, 0.4, 0.5, 0.6) were prepared by solid-state reactions with using a sol-gel produced precursor~\cite{Zhang1994,Wang2024}. All the samples with $x\leq0.5$ remain to be monophasic, as checked by powder X-ray diffractions (XRD), yet impurity phases appear for the sample of $x=$ 0.6. This means that the Al solubility limit is about $x=$ 0.5 under current synthesis conditions. The as-prepared monophasic samples were annealed at 500~$^{\circ}$C under $\sim$10 MPa oxygen atmosphere, the outcome of which is called post-annealed samples. Details of the sample preparation and treatment are provided in Methods. The chemical composition and homogeneity were checked by energy dispersive X-ray (EDX) spectroscopy. The results show homogeneous composition that is almost identical to the nominal stoichiometry within the measurement uncertainty for cations including the dopant Al [Figure S1 in the Supporting Information (SI)].

\begin{figure}[h]
	\centering
	\includegraphics[width=0.9\textwidth]{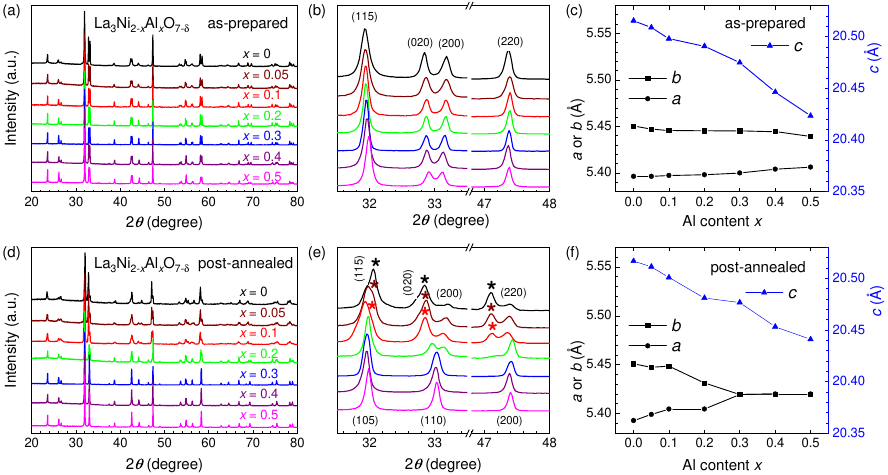}
	\caption{Characterization of as-prepared (a-c) and post-annealed (d-f) samples of La$_3$Ni$_{2-x}$Al$_x$O$_{7-\delta}$ ($0\leq x\leq0.5$) by powder X-ray diffractions. In panel (e), the peaks marked by asterisks are attributed to a new phase formed during the annealing (see the text and SI). Plots (c) and (f) show lattice parameters as functions of Al content $x$. The error bars are smaller than the symbols.
		\label{XRD1}}
\end{figure}
Figures~\ref{XRD1}(a-c) show the XRD results for the as-prepared samples of La$_3$Ni$_{2-x}$Al$_x$O$_{7-\delta}$ ($0\leq x\leq0.5$). All the XRD peaks can be well indexed with the $Amam$ structure~\cite{Ling2000}, and no secondary phase is detectable. The lattice parameters are calculated by least-squares fit of the reflections indexed, which are plotted as functions of Al content $x$ in Figure~\ref{XRD1}(c). One sees that the $c$ axis is reduced significantly by the Al doping, which is attributed to the smaller size of Al$^{3+}$ ions compared with Ni$^{2.5+}$ ions. Meanwhile, the $a$ and $b$ axes only change slightly, and they tend to get closer with the Al doping. The orthorhombicity, defined by $\epsilon = 2(b-a)/(b+a)\times100\%$, decreases from 1.00$\%$ for $x$ = 0 to 0.61$\%$ for $x$ = 0.5, which aligns with the increase of the tolerance factor. Nevertheless, our aim for the tetragonal structure is not achieved yet in those as-prepared samples.

In order to obtain the target tetragonal phase, we tried to increase the oxygen content by annealing the samples under high oxygen pressure (see Methods for the details). 
The result is presented in Figures~\ref{XRD1}(d-f). For the non-doped sample of $x=$ 0, first of all, the intensity of (220) reflection at 2$\theta\approx47.4^{\circ}$ is obviously reduced [Figure~\ref{XRD1}(e)], and an additional peak nearby (marked with asterisks) appears, indicating formation of a new phase. A similar observation was also reported very recently~\cite{327damage}. The new phase (here we call it $\beta$ phase) can be best fitted with a tetragonal $I4/mmm$ unit cell: $a = 3.8552(1)$~\AA~ and $c = 20.2120(8)$ \AA, both of which are very close to those of the tetragonal La$_3$Ni$_2$O$_7$ phase reported very recently~\cite{La327.CXH,LIU2025125528}. The weight fraction of the $\beta$ phase is 75.4\%, as opposed to 24.6\% for the remaining $\alpha$ phase. The $\alpha$ phase has the normal orthorhombic bilayer $Amam$ structure, with almost identical lattice parameters of the as-prepared sample (Figure S2 in SI). For the sample of $x=0.1$, the two-phase Rietveld analysis yields similar lattice parameters for the $\alpha$ and $\beta$ phases, yet shows a reversed result (80.5\% and 19.5\%, respectively) for their weight percentages. The fraction of new phase for $x=0.2$ is near the XRD detection limit, say 1-5 wt.\%, because only discernibly small signal from the new phase is seen. 

\begin{figure}[h]
	\centering
	\includegraphics[width=0.9\textwidth]{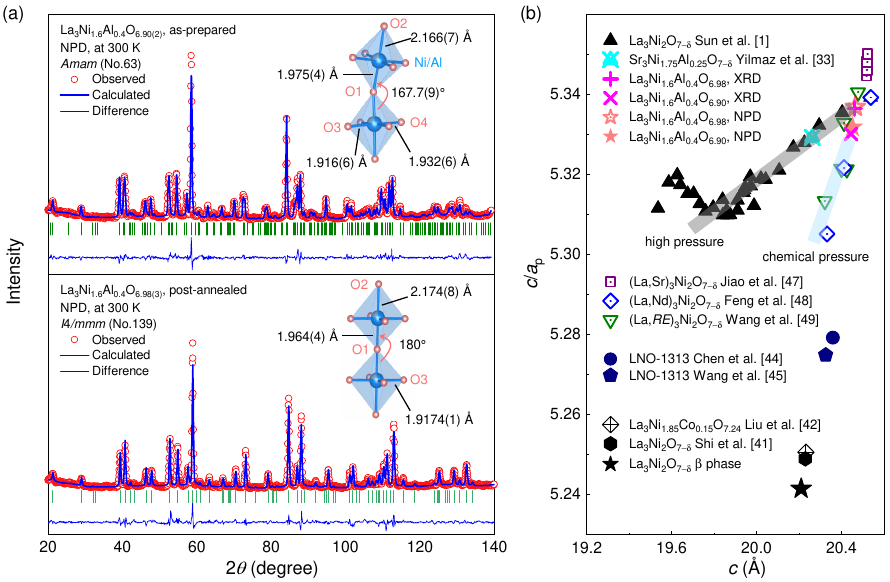}
	\caption{(a) Rietveld refinement profiles of the neutron diffractions for the as-prepared (top) and post-annealed (bottom) samples of La$_3$Ni$_{1.6}$Al$_{0.4}$O$_{7-\delta}$. The insets show bond distances and angles in the NiO$_6$ bilayers. (b) The axial ratios $c/a_\mathrm{p}$ as functions of $c$ for the bilayer and monolayer-trilayer (1313) nickelates. The simple perovskite unit, $a_\mathrm{p}=\frac{1}{2}\sqrt{a^2+b^2}$, is employed to effectively compare the tetragonal phases with orthorhombic ones.
		\label{XRD2}}
\end{figure}
For the heavily-doped samples of $0.3\leq x\leq 0.5$, interestingly, the original (020) and (200) reflections (on the basis of an orthorhombic lattice) merge into a single peak [(110) reflection in the tetragonal lattice, Figure~\ref{XRD1}(e)], suggesting formation of the target tetragonal structure. Note that this tetragonal phase is different from the $\beta$ phase mentioned above (see below). Figure~\ref{XRD1}(f) displays the lattice parameters as functions of Al doping, which clearly shows the structural evolution from orthorhombic to tetragonal at $x\approx0.3$. The result contrasts with that of the as-prepared samples [Figure~\ref{XRD1}(c)], underscoring the importance of annealing under high oxygen pressure. 

To determine the oxygen stoichiometry, we performed thermogravimetric analysis (TGA) measurements on selected representative samples [shown in Figure S3 in SI]. The results demonstrate a clear enhancement in thermal stability with increasing Al doping concentration, as evidenced by reduced weight loss percentages [5.82$\%$ ($x=$ 0) to 0.58$\%$ ($x=$ 0.4) for as-prepared samples, 5.98$\%$ ($x=$ 0) to 0.89$\%$ ($x=$ 0.4)  for post-annealed samples]. The result also shows that post-annealed samples consistently exhibit higher oxygen content than as-prepared counterparts. For La$_3$Ni$_{1.9}$Al$_{0.1}$O$_{7-\delta}$ samples, analysis of the weight-loss plateaus corresponding to La$_3$Ni$_{1.9}$Al$_{0.1}$O$_{6.45}$~\cite{2024AMI.Cava} yields oxygen contents of about 6.87 (as-prepared) and 7.00 (post-annealed). The improved oxygen content and thermal stability in Al-doped systems suggest that Al incorporation may effectively stabilize the oxygen sublattice during thermal treatment.

The crystallographic data of the tetragonal phase were obtained by the Rietveld analysis based on the XRD slow-scan data. The XRD refinement profiles for the typical samples of $x=0.4$ are displayed in Figure S3, and the resulted data are presented in Table S1, both of which are deposited in the SI. Note that the $c$-axis value of 20.4620(4) \r{A} of the tetragonal sample of $x=0.4$ is even obviously larger than that of the $\beta$ phase of $x=0$. It is also larger than that of the so-called 1313 phase (another polymorph of La$_3$Ni$_2$O$_7$, crystallizing in a hybrid monolayer-trilayer structure~\cite{1313.jacs,1313.XWW,1313.prl}). To highlight the differences, in Figure~\ref{XRD2}(b) we plot the axial ratios, $c/a_\mathrm{p}$, where $a_\mathrm{p}=\frac{1}{2}\sqrt{a^2+b^2}$ in the case of an orthorhombic phase, as functions of $c$ axis for various nominal ``327'' phases~\cite{Jiao2024.PC,Feng2024.PRB,wang2025.npj,1313.jacs,1313.XWW}. As is seen, the Al-doped tetragonal or orthorhombic phase has a similar axial ratio with those of the superconducting La$_3$Ni$_2$O$_7$ under high pressures~\cite{Sun2023}, and with those of La-site doped La$_3$Ni$_2$O$_7$~\cite{Jiao2024.PC,Feng2024.PRB,wang2025.npj}. However, the $c/a_\mathrm{p}$ value of the $\beta$ phase is the smallest among the series. Thus the non-doped tetragonal $\beta$ phase might possess substantially different structures which lead to absence of superconductivity even under high pressures~\cite{La327.CXH}.
\begin{table}[htbp]
	\caption{Crystallographic data of the as-prepared (orthorhombic phase) and post-annealed (tetragonal phase) La$_3$Ni$_{1.6}$Al$_{0.4}$O$_{7-\delta}$ from neutron powder diffractions.}
	\label{NPD}
	\begin{tabular}{ccccccc}
		\toprule
		\multicolumn{3}{c}{orthorhombic phase}  &\multicolumn{4}{c}{La$_3$Ni$_{1.6}$Al$_{0.4}$O$_{6.90}$}   \\
		\multicolumn{3}{c}{space group}    &\multicolumn{4}{c}{ $Amam$ (No. 63)}    \\
		\multicolumn{3}{c}{$a$ (\r{A})}         &\multicolumn{4}{c}{5.4055(1)}   \\
		\multicolumn{3}{c}{$b$ (\r{A})}         &\multicolumn{4}{c}{5.4438(1)} \\
		\multicolumn{3}{c}{$c$ (\r{A})}         &\multicolumn{4}{c}{20.4515(5)}   \\
		\multicolumn{3}{c}{$R_\mathrm{wp}$  (\%)}             &\multicolumn{4}{c}{6.18} \\
		\multicolumn{3}{c}{$S$}             &\multicolumn{4}{c}{1.65} \\
		\midrule
		atom & site & $x$     & $y$       & $z$        &  occ.  &  $U_{iso}$                    \\
		% \hline
		La(1)    &  4$c$  & 0.25     & 0.240(2)    & 0.5           & 1.00        & 0.009(1)               \\
		La(2)    &  8$g$  & 0.25     & 0.243(2)    & 0.3191(1)   & 1.00        & 0.009(1)                     \\
		Ni/Al    &  8$g$  & 0.25     & 0.247(2)    & 0.0960(2)   & 0.80/0.20   & 0.003(1)                \\
		O(1)     &  4$c$  & 0.25     & 0.208(2)    & 0             & 0.90(2)     & 0.002(3)                \\
		O(2)     &  8$g$  & 0.25     & 0.283(1)    & 0.2015(3)   & 1.00        & 0.010(1)               \\
		O(3)     &  8$e$  & 0.5      & 0           & 0.1049(3)   & 1.00        & 0.015(2)               \\
		O(4)     &  8$e$  & 0        & 0.5         & 0.0901(3)   & 1.00        & 0.001(2)                     \\
		\midrule\midrule
		\multicolumn{3}{c}{tetragonal phase}    &\multicolumn{4}{c}{La$_3$Ni$_{1.6}$Al$_{0.4}$O$_{6.98}$}    \\
		\multicolumn{3}{c}{space group}    &\multicolumn{4}{c}{ $I4/mmm$ (No. 139)}    \\
		\multicolumn{3}{c}{$a$ (\r{A})}         &\multicolumn{4}{c}{3.83466(6)}   \\
		\multicolumn{3}{c}{$c$ (\r{A})}         &\multicolumn{4}{c}{20.4648(5)}   \\
		\multicolumn{3}{c}{$R_\mathrm{wp}$  (\%)}             &\multicolumn{4}{c}{7.83} \\
		\multicolumn{3}{c}{$S$}             &\multicolumn{4}{c}{2.08} \\
		\midrule
		atom       & site & $x$     & $y$       & $z$        &  occ.  &  $U_{iso}$                    \\
		% \hline
		La(1)      &  4$e$  & 0     & 0       & 0.3184(2)     & 1.00     & 0.009(1)             \\
		La(2)      &  2$b$  & 0     & 0       & 0.5           & 1.00     & 0.009(1)                    \\
		Ni/Al      &  4$e$  & 0     & 0       & 0.0960(2)     & 0.80/0.20     & 0.010(1)                 \\
		O(1)       &  2$a$  & 0     & 0       & 0             & 0.98(3)  & 0.041(4)                \\
		O(2)       &  4$e$  & 0     & 0       & 0.2022(3)     & 1.00     & 0.025(1)               \\
		O(3)       &  8$g$  & 0     & 0.5     & 0.0950(2)   & 1.00     & 0.017(1)               \\
		\bottomrule
	\end{tabular}
\end{table}

To more precisely determine the oxygen positions and occupancies, we conducted neutron powder diffraction (NPD) measurements on both the as-prepared and the post-annealed samples of $x=0.4$, as shown in Figure~\ref{XRD2}(a). The refined results, summarized in Table~\ref{NPD}, reveal that the oxygen occupancies at the O1 site ($i.e.$, the apical site between NiO$_2$ planes) increases from  0.90(2) in the as-prepared to 0.98(3) in the post-annealed samples. Notably, refinement of the post-annealed sample using a model that includes interstitial oxygen (adopting the configuration observed in La$_2$NiO$_{4+\delta}$ where excess oxygen resides between LaO layers~\cite{JPCM1991,B605886H}) yields a higher $R_\mathrm{wp}$ value (8.8 $\%$), suggesting absence of interstitial oxygen here. As illustrated in the insets of Figure~\ref{XRD2}(a), the Ni$-$O1$-$Ni bond angle changes expectedly from 167.7(9)$^{\circ}$ (in the $Amam$ phase) to 180$^{\circ}$ (in the $I4/mmm$ phase), expectedly increasing the interlayer coupling. At the same time, the Ni$-$O3$-$Ni bond also becomes straight in the tetragonal phase, which enhances the hybridization between Ni-$3d_{x^2-y^2}$ and O-2$p_{x/y}$. Albeit with the Al doping, contrastingly, the Ni$-$O bond distances in the NiO$_6$ octahedra do not change significantly. Based on the bond distances, the bond valence sums (BVS)~\cite{BVS} calculated (without consideration of the Al doping) are 2.55 and 2.62, respectively, for the as-prepared and post-annealed samples, consistent with the increase of oxygen content. Finally, the $c/a_\mathrm{p}$ value increases a little with post annealing, in sharp contrast with the very low value of the $\beta$ phase in the post-annealed La$_3$Ni$_2$O$_7$. The realization of the tetragonal phase at ambient pressure here can be basically understood in terms of the concept of Goldschmidt tolerance factor $\tau$. First, $\tau$ is increased by the Al doping because of the difference in the ionic radii of Ni$^{2.5+}$ and Al$^{3+}$. Second, $\tau$ is further increased by the increase of oxygen occupancy at the O1 site. Thus, the Al doping and oxygenation have a synergistic effect on $\tau$, which makes the tetragonal phase finally stabilized. 

\begin{figure}[h]
	\centering
	\includegraphics[width=0.9\textwidth]{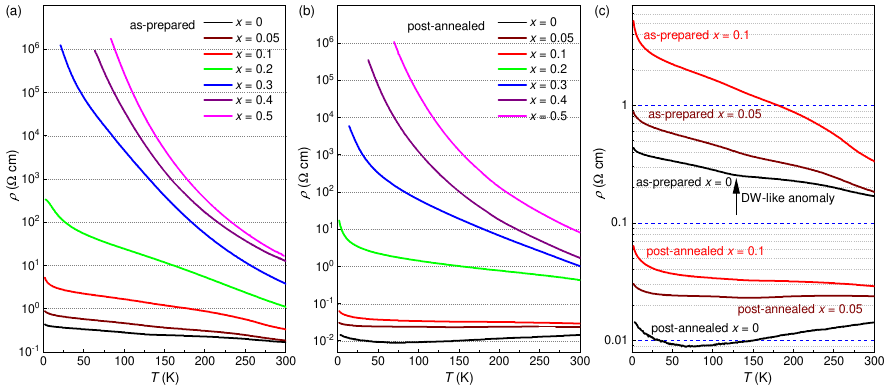}
	\caption{Temperature dependence of electrical resistivity of the as-prepared (a) and post-annealed (b) samples of La$_3$Ni$_{2-x}$Al$_x$O$_{7-\delta}$. The effect of post annealing is highlighted in panel (c) for $x=$ 0, 0.05 and 0.1.
		\label{RT}}
\end{figure}
Electrical resistivity measurements at ambient pressure on La$_3$Ni$_{2-x}$Al$_x$O$_{7-\delta}$ demonstrate absence of superconducivity in all samples, as shown in Figure~\ref{RT}. One sees that the resistivity increases rapidly with Al doping, and the Al-doped samples show semiconducting or insulating behaviors. The result suggests strong carrier localization induced by Al-doping disorder. The low-temperature resistivity basically follows variable-range hopping (VRH) formula $\rho(T) = \rho_{0}$exp($T_\mathrm{0}$/$T$)$^{1/D}$, where $D$ refers to the dimension in the electron hopping [Figure S4 in SI]~\cite{shklovskii2013electronic}. The two-dimensional VRH better fits the data, suggesting that disorder effects are confined to the NiO$_2$ planes. One also note that the resistivity of the post-annealed samples is systematically reduced, which inversely implies that the oxygen vacancy in the as-prepared samples also contributes the carrier localization~\cite{Taniguchi1995}. For the as-prepared undoped sample, an anomaly at $\sim$125 K can be detected, similar to the earlier reports in La$_3$Ni$_{2}$O$_{6.92}$ polycrystalline sample~\cite{Zhang1994,Taniguchi1995}. The anomaly has been recently identified as the density-wave (DW) transition~\cite{Sun2023,Liu2023.SCP.WangM}. This DW-like transition is smeared out at $x=0.1$, and becomes indiscernible for $x\geq0.2$. The result indicates that the Al doping destroys the DW order either~\cite{periyasamy2021}. For the post-annealed samples, no signature of DW transition is present even in the undoped parent compound, primarily due to phase transformation into the $\beta$ phase, similar to the very recent report~\cite{La327.CXH}. 

\begin{figure}
	\centering
	\includegraphics[width=0.8\textwidth]{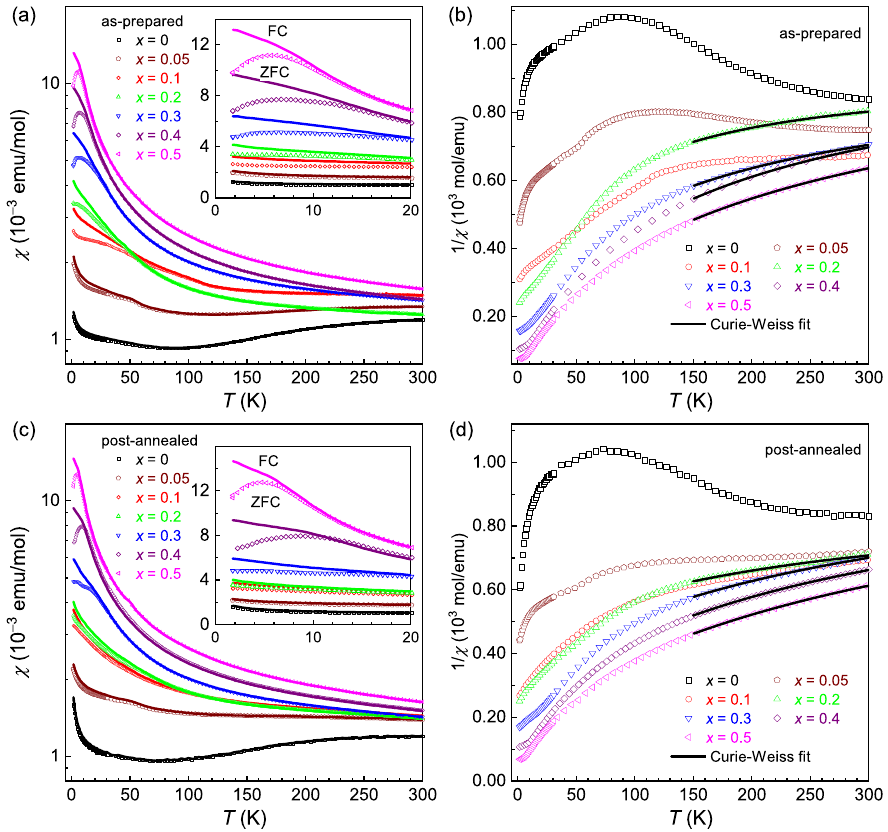}
	\caption{Magnetic susceptibility data for the as-prepared (a,b) and post-annealed (c,d) samples of La$_3$Ni$_{2-x}$Al$_x$O$_{7-\delta}$. In panels (a) and (c), a logarithmic scale is used to distinctly show the lower-value data. FC and ZFC denote field cooling and zero field cooling, respectively.
		\label{MT}}
\end{figure}
Figure~\ref{MT} shows the magnetic susceptibility data measured under a magnetic field of $\mu_0H = 0.1$~T for all samples of La$_3$Ni$_{2-x}$Al$_x$O$_{7-\delta}$. The undoped samples exhibit relatively low values of magnetic susceptibility ($\sim$0.001 emu/mol) with a weak temperature dependence. At high temperatures above $\sim$100 K, the susceptibility decreases with decreasing temperature, suggesting existence of antiferromagnetic correlations. Upon Al doping, the magnetic susceptibility increases systematically, and the high-temperature $\chi(T)$ data for $x>$ 0.1 basically follows the extended Curie-Weiss formula, $\chi= \chi_0+C/(T+\theta_\mathrm{W})$, where $\chi_0$, $C$, and $\theta_\mathrm{W}$ represent the temperature-independent term, the Curie constant, and the paramagnetic Curie-Weiss temperature, respectively. The data fitting in the temperature range of 150-300~K [Figures~\ref{MT}(b) and 4(d)] yields the three parameters, which are presented in Table S2 in the SI. The $\chi_0$ values are all around 0.001 emu/mol, basically identical to the $\chi$ value of the undoped samples. The Curie constant increases progressively with Al doping, and the derived effective magnetic moments($\mu_\mathrm{eff}$) increase from 0.48 ($x=$ 0.2) to 1.01 ($x=$ 0.5) $\mu_{\rm B}$/Ni for the as-prepared samples. A similar trend is also observed for the post-annealed samples with $x>0.2$. At low temperatures, one sees that the FC and ZFC data bifurcate for Al-doped samples, suggesting freezing of localized spins. Evidenced of the spin freezing is also given by the magnetic hysteresis at the lowest temperature measured (Figure S5 in the SI). While oxygen annealing increases the absolute susceptibility values very slightly, the magnetic parameters remain nearly unchanged, demonstrating that Al doping rather than oxygen content dominates the magnetic modifications, mirroring observations in Zn-doped cuprates where nonmagnetic impurities induce moments independent of oxygen-content variations~\cite{Mahajan1994.PRL.Zn-doping,1995PhysRevB.Zagolaev}.

\begin{figure}
	\centering
	\includegraphics[width=0.9\textwidth]{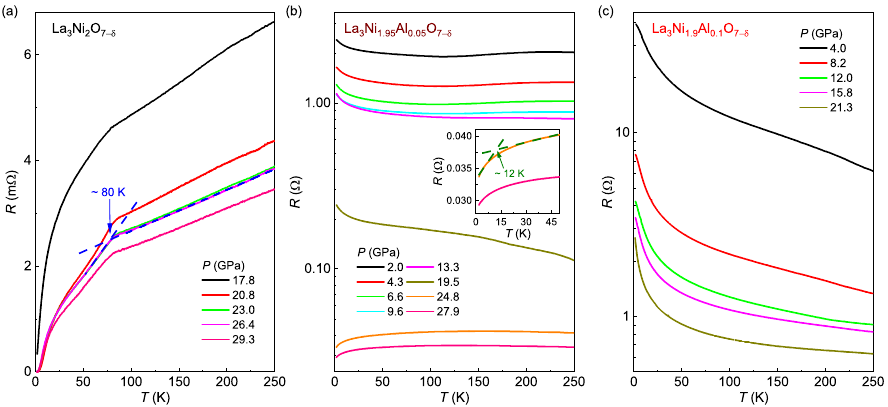}
	\caption{Temperature dependence of resistance under pressures for La$_3$Ni$_2$O$_{7-\delta}$ (a),  La$_3$Ni$_{1.95}$Al$_{0.05}$O$_{7-\delta}$ (b) and La$_3$Ni$_{1.9}$Al$_{0.1}$O$_{7-\delta}$ (c). $T_\mathrm{c}$ is determined as the interception between two linear extrapolations below and above the superconducting transition. The inset in panel (b) shows an enlargement of the $R(T)$ for $x =$ 0.05 sample at 24.8 and 27.9~GPa below 50 K. phases with orthorhombic ones.
		\label{HP}}
\end{figure}
To investigate the potential pressure-induced superconductivity in Al-doped La$_3$Ni$_2$O$_{7-\delta}$, we performed temperature-dependent resistance measurements on the undoped sample La$_3$Ni$_2$O$_{7-\delta}$ (pure orthorhombic phase after oxygen-annealing at 400~$^\circ$C) and post-annealed samples ($x$ = 0.05, 0.1, and 0.4) under pressures. As shown in Figure~\ref{HP}(a), the undoped sample displays a superconducting transition at $T_\mathrm{c} \approx$ 80~K, consistent with previous reports~\cite{Sun2023,2024PRX.ChengJG,2024NP.YuanHQ}. The observed broad transition might arise from sample inhomogeneity and/or weak links between crystalline grains. The $x =$ 0.05 sample [Figure~\ref{HP}(b)] maintains semiconducting behavior at low pressures but shows a pronounced resistance drop below 12~K at 24.8~GPa. This drop is probably associated with the emergence of a superconducting transition. In contrast, the $x =$ 0.1 sample [Figure~\ref{HP}(c)] and tetragonal phase $x$ = 0.4 sample [Figure S7 in SI] keep semiconducting/insulating behavior within measured pressure range and have large resistance values even over 20~GPa, implying stronger carrier localization with increased Al doping. Furthermore, the orthorhombic $x = 0.05$ sample annealed at the lower temperature of 400~$^\circ$C remains insulating at 24.7~GPa [Figure S8 in SI], likely due to insufficient oxygen content. These results indicate that a low Al substitution at $\sim 2.5\%$ can effectively destroy superconductivity in this system.

\begin{figure}
	\centering
	\includegraphics[width=0.7\textwidth]{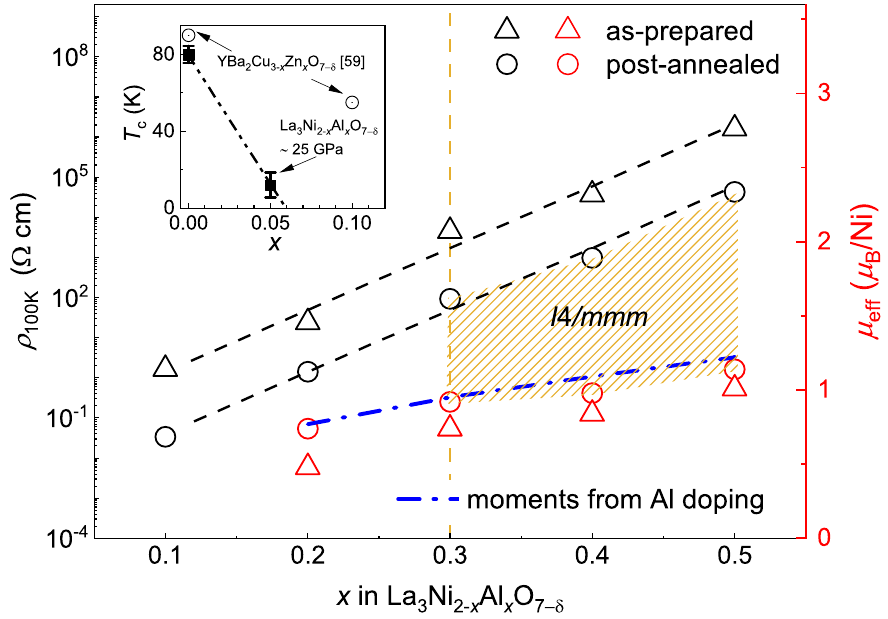}
	\caption{Electrical resistivity at 100 K (left axis) and effective magnetic moments (right axis) as functions of Al doping in La$_3$Ni$_{2-x}$Al$_x$O$_{7-\delta}$. The black dashed lines are guides to the eye, and the dash-dotted line represents the expected effective moments induced by Al doping. The yellow trapezoidal area merely indicates the experimental parameter range (0.3 $\leq x \leq$ 0.5) where the $I4/mmm$ tetragonal phase was observed in this study. The inset demonstrates the relationship between $T_\mathrm{c}$ and doping content in La$_3$Ni$_{2-x}$Al$_x$O$_{7-\delta}$ at $\sim$ 25 GPa and YBa$_2$Cu$_{3-x}$Zn$_x$O$_{7-\delta}$ at ambient pressure~\cite{tarascon1988structural}.
		\label{localization}}
\end{figure}

Figure~\ref{localization} plots the ambient-pressure electrical resistivity at 100 K, $\rho_{\mathrm{100K}}$, and the effective magnetic moments derived above as functions of Al content. The $\rho_{\mathrm{100K}}$ value increases almost exponentially with the Al doping, independent of the oxygenation by post annealing (although the resistivities are nearly two orders of magnitude reduced) and the crystal-structure transformation. The result indicates that the Al dopant dominates the carrier localization. Remarkably, such a localization can be weakened by the $A$-site chemical substitution, as suggested from a latest report~\cite{yilmaz2025}. Note that free local moments are absent in the undoped samples, while the Al-doped samples do carry local magnetic moments. The effective magnetic moments originate from Al doping, similar to the case in cuprates~\cite{Mahajan1994.PRL.Zn-doping,1995PhysRevB.Zagolaev,Ishida1996.PRL.Al-doping}. As shown in Figure~\ref{localization}, the calculated values of effective moments, assuming that each Al atom induces a spin-1/2 localized moment, align closely with the experimental ones. This supports the interpretation that Al$^{3+}$ locally disrupts the magnetic order in the NiO$_{2}$ planes. The inset of Figure~\ref{localization} shows that $T_\mathrm{c}$ decreases rapidly as Al doping increases in La$_3$Ni$_{2-x}$Al$_x$O$_{7-\delta}$, similar to the effect of Zn doping in YBa$_2$Cu$_{3-x}$Zn$_x$O$_{7-\delta}$~\cite{tarascon1988structural}. Given that nonmagnetic scattering brgaieaks Cooper pairs and significantly suppresses $T_\mathrm{c}$ in unconventional superconductors~\cite{anderson1959theory,millis1988inelastic}, the dramatic reduction of $T_c$ by a small amount of nonmagnetic impurity provides strong evidence for unconventional superconductivity in La$_3$Ni$_2$O$_{7-\delta}$.

In summary, we have successfully realized the tetragonal phase of bilayer nickelate at ambient pressure by Al doping together with post annealing in oxygen atmosphere ($P_{\mathrm{O}_2}\approx$ 10 MPa). The stabilization of the tetragonal structure can be basically understood with the concept of Goldschmidt tolerance factor. We also found that the post-annealed undoped sample partially transforms into another tetragonal phase--the $\beta$ phase. This $\beta$ phase is significantly different from the Al-doped tetragonal one in the light of the axial ratio $c/a_\mathrm{p}$. Note that the $c/a_\mathrm{p}$ value of the latter is similar to that of the superconducting bilayer nickelate under high pressures. Thus, our work provides an initial step towards the ultimate realization of bulk HTSC at ambient pressure.

The aluminum doping also serves as a probe in the system, which significantly alters the physical properties. Even at a low doping level, the system evolves from a bad metal to a semiconductor, and significantly suppresses pressure-induced superconductivity. The resistivity increases almost exponentially with the Al doping, indicating that the Al doping induces strong disorder. Notably, the Al doping also induces local magnetic moments, a phenomenon reminiscent of nonmagnetic doping effect in cuprate superconductors. It would be worthwhile to dope with an element in proximity to nickel to lower the localization effect in the future.   

\section{METHODS}
Polycrystalline samples of La$_3$Ni$_{2-x}$Al$_x$O$_{7-\delta}$ (0 $\leq x\leq$ 0.6) were synthesized by solid-state reactions using a sol-gel produced precursor.~\cite{Zhang1994,Wang2024} Stoichiometric mixture of the source materials, La(NO$_3$)$_3$$\cdot$6H$_2$O (99.9\% Aladdin), Ni(NO$_3$)$_2$$\cdot$6H$_2$O (99.99\% Aladdin), and Al(NO$_3$)$_3$$\cdot$9H$_2$O (99.99\% Aladdin), were dissolved in the glycol and deionized water with addition of appropriate amount of citric acid. The mixed solution was continuously stirred in a water bath (90~$^\circ$C) for 4 h, and homogeneous green gel resulted. The gel was slowly heated to 500~$^\circ$C in air, and then further heated to 800~$^\circ$C, holding for 10 h to eliminate organic components. The resulted precursor was ground and pressed into pellets, and the pellets were sintered in air at 1080-1150~$^\circ$C for 50 h, which produces single-phase samples of La$_3$Ni$_{2-x}$Al$_x$O$_{7-\delta}$. In order to obtain the tetragonal phase, the as-prepared samples were post annealed at 500~$^\circ$C for 30 h in 10 MPa oxygen atmosphere. %The range was selected as it was sufficient to stabilize the orthorhombic phase and approached the Al solubility window in $Amam$ (No. 63) space group.

Powder XRD data were collected on a PANalytical diffractometer (Empyrean Series 2) with Cu-$K_{\alpha1}$ radiation. TGA measurements were accomplished in HQT-3, using a 10$\%$ H$_2$/Ar gas flow of 3 mL/min with a 20~$^\circ$C/min rate up to 740~$^\circ$C. The NPD measurements were carried out on the high-resolution neutron diffractometer at the Key Laboratory of Neutron Physics, Institute of Nuclear Physics and Chemistry, China Academy of Engineering Physics. The
wavelength of the neutron was $\lambda$ = 1.8846 \r{A}. The crystal structure was refined by Rietveld analysis using the GSAS-\uppercase\expandafter{\romannumeral2} package~\cite{toby2013gsas}.
The chemical composition was determined using EDX spectroscopy on a scanning electron microscope (Hitachi S-3700N) equipped with Oxford Instruments X-Max spectrometer. Electrical resistivity was measured on a Quantum Design Physical Property Measurement System (PPMS-9) using standard four-electrodes method. The magnetic properties were measured on a Quantum Design Magnetic Property Measurement System (MPMS3). Resistance measurements under pressures were performed in a diamond anvil cell (DAC) using Daphne oil 7373 as the presssure-transmitting medium~\cite{2024NP.YuanHQ}, pressure is determined at room temperature using the ruby fluorescence method~\cite{mao1986calibration}.

\bmhead{Data availability}
The data that support the findings of this study are available from the corresponding author upon reasonable request.

\backmatter

%\bmhead{Supplementary information}

%If your article has accompanying supplementary file/s please state so here. 

%Authors reporting data from electrophoretic gels and blots should supply the full unprocessed scans for key as part of their Supplementary information. This may be requested by the editorial team/s if it is missing.

%Please refer to Journal-level guidance for any specific requirements.

\bmhead{Acknowledgements}

We thank Wu-Zhang Yang and Zhi Ren for their help in the electrical resistivity measurement. This work was supported by the National Natural Science Foundation of China (12494593, 12494592 and 12034017), the National Key Research and Development Program of China (2022YFA1403202, 2022YFA1402200 and 2023YFA1406101) and the CAS Superconducting Research Project under Grant No. [SCZX-0101].

\bmhead{Competing interests}

The authors declare no competing interests.

\bmhead{Author contributions}

G.H.C designed the project; J.Y.L. synthesized the materials and measured transport and magnetic properties; K.X.Y. and Y.N.Z. conducted the high-pressure resistance measurements; H.L. and B.J.L. measured the NPD data; J.X.L. measured the TGA data; J.Y.L. and Y.Q.L. carried out the structural characterizations (XRD and NPD) and performed data analysis with the support of G.H.C, H.Q.Y., K.X.Y., X.Y.Z., J.X.L. and Y.N.Z.; G.H.C., J.Y.L. and Y.Q.L. wrote the paper with inputs from all coauthors.

\clearpage
\begin{appendices}
\setcounter{figure}{0}
\setcounter{table}{0}
\renewcommand{\thefigure}{S\arabic{figure}}
\renewcommand{\thetable}{S\arabic{table}}

\section{Supplemental Information: Stabilization of Tetragonal Phase and Aluminum-Doping Effect in a Bilayer Nickelate}\label{secA1}

Contents

Figure S1. Energy dispersive X-ray (EDX) spectroscopy for La$_3$Ni$_{1.6}$Al$_{0.4}$O$_{7-\delta}$.

Figure S2. XRD Rietveld refinements for post-annealed La$_3$Ni$_{2}$O$_{7-\delta}$ and La$_3$Ni$_{1.9}$Al$_{0.1}$O$_{7-\delta}$.

Figure S3. Thermogravimetric curves for representative as-prepared and post-annealed La$_3$Ni$_{2-x}$Al$_x$O$_{7-\delta}$.

Figure S4. XRD Rietveld refinements for as-prepared and post-annealed La$_3$Ni$_{1.6}$Al$_{0.4}$O$_{7-\delta}$.

Figure S5. The linear fits of ln$\rho$ to $T^{-1/3}$ and $T^{-1/4}$ for La$_3$Ni$_{2-x}$Al$_x$O$_{7-\delta}$ samples.

Figure S6. Isothermal magnetization for La$_3$Ni$_{2-x}$Al$_x$O$_{7-\delta}$ samples.

Figure S7. Temperature dependence of resistance under pressures for post-annealed La$_3$Ni$_{2-x}$Al$_x$O$_{7-\delta}$. 

Figure S8. Temperature dependence of resistance under pressures for La$_3$Ni$_{1.95}$Al$_{0.05}$O$_{7-\delta}$ annealed at 400~$^\circ$C in oxygen.

Table S1. Crystallographic data of La$_3$Ni$_{1.6}$Al$_{0.4}$O$_{7-\delta}$ from XRD.

Table S2. The Curie-Weiss fit parameters for La$_3$Ni$_{2-x}$Al$_x$O$_{7-\delta}$ (0.2 $\leq x\leq$ 0.5).

\clearpage
\begin{figure*}[htbp]
	\includegraphics[width=10cm]{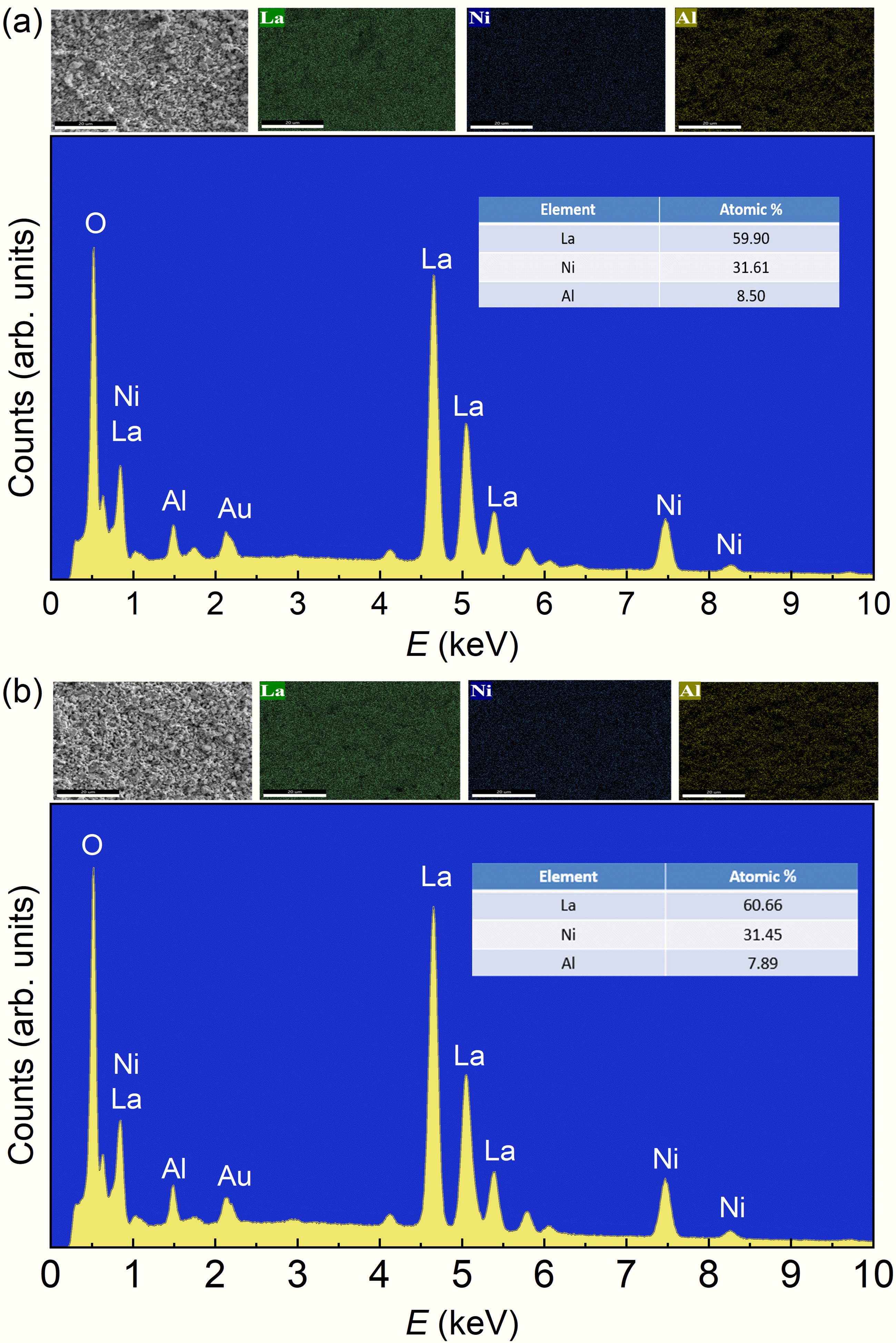}
    \centering
	\caption{Verification of the chemical compositions of the as-prepared (a) and post-annealed (b) samples of La$_3$Ni$_{1.6}$Al$_{0.4}$O$_{7-\delta}$ by energy dispersive X-ray spectroscopy. The atomic ratios of cations are La: Ni: Al = 3: 1.58: 0.43 and 3: 1.56: 0.39, respectively, for the as-prepared and post-annealed samples. The oxygen content is not analyzed here because of the inherent large measurement uncertainty. Note that the surface of the sample was sprayed with gold, in order to avoid the accumulation of electric charge.}
	\label{figS1}
\end{figure*}

\begin{figure*}[htbp]
	\includegraphics[width=10cm]{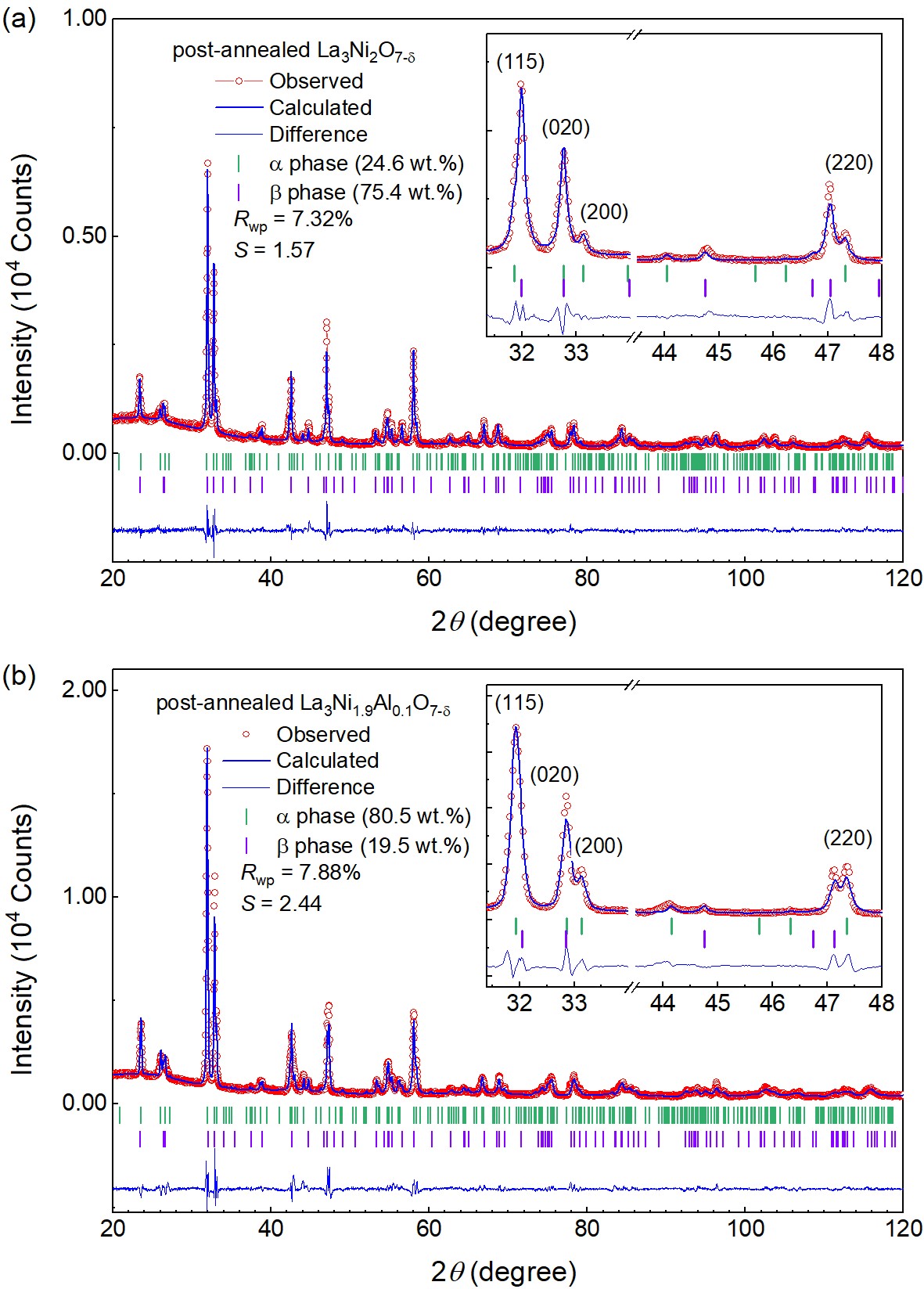}
    \centering
	\caption{Two-phase Rietveld refinement profiles based on powder X-ray diffraction data for post-annealed La$_3$Ni$_{2}$O$_{7-\delta}$ (a) and La$_3$Ni$_{1.9}$Al$_{0.1}$O$_{7-\delta}$ (b). The $\alpha$ phase has the normal orthorhombic bilayer structure ($Amam$), while the $\beta$ phase is modelled with K$_3$Ni$_{2}$F$_{7}$-type structure ($I$4$/mmm$). The insets show some of the main reflections. For non-doped La$_3$Ni$_{2}$O$_{7-\delta}$, the refined lattice parameters are, $a$ = 5.3931(4)~\r{A}, $b$ = 5.4509(6)~\r{A}, $c$ = 20.5167(15)~\r{A} for the La$_3$Ni$_{2}$O$_{7-\delta}$ $\alpha$ phase (24.6~wt.$\%$); and $a$ = 3.8552(1)~\r{A}, $c$ = 20.2120(8)~\r{A} for the $\beta$ phase (75.4 wt.$\%$). In the case of the Al-doped La$_3$Ni$_{1.9}$Al$_{0.1}$O$_{7-\delta}$, the cell parameters are $a$ = 5.4049(2)~\r{A}, $b$ = 5.4487(2)~\r{A}, $c$ = 20.5009(10)~\r{A} for the $\alpha$ phase (80.5 wt.$\%$); and $a$ = 3.8543(2)~\r{A}, $c$ = 20.2418(17)~\r{A} for the $\beta$ phase (19.5~wt.$\%$).}
	\label{figS2}
\end{figure*}

\begin{figure*}[htbp]
	\includegraphics[width=10cm]{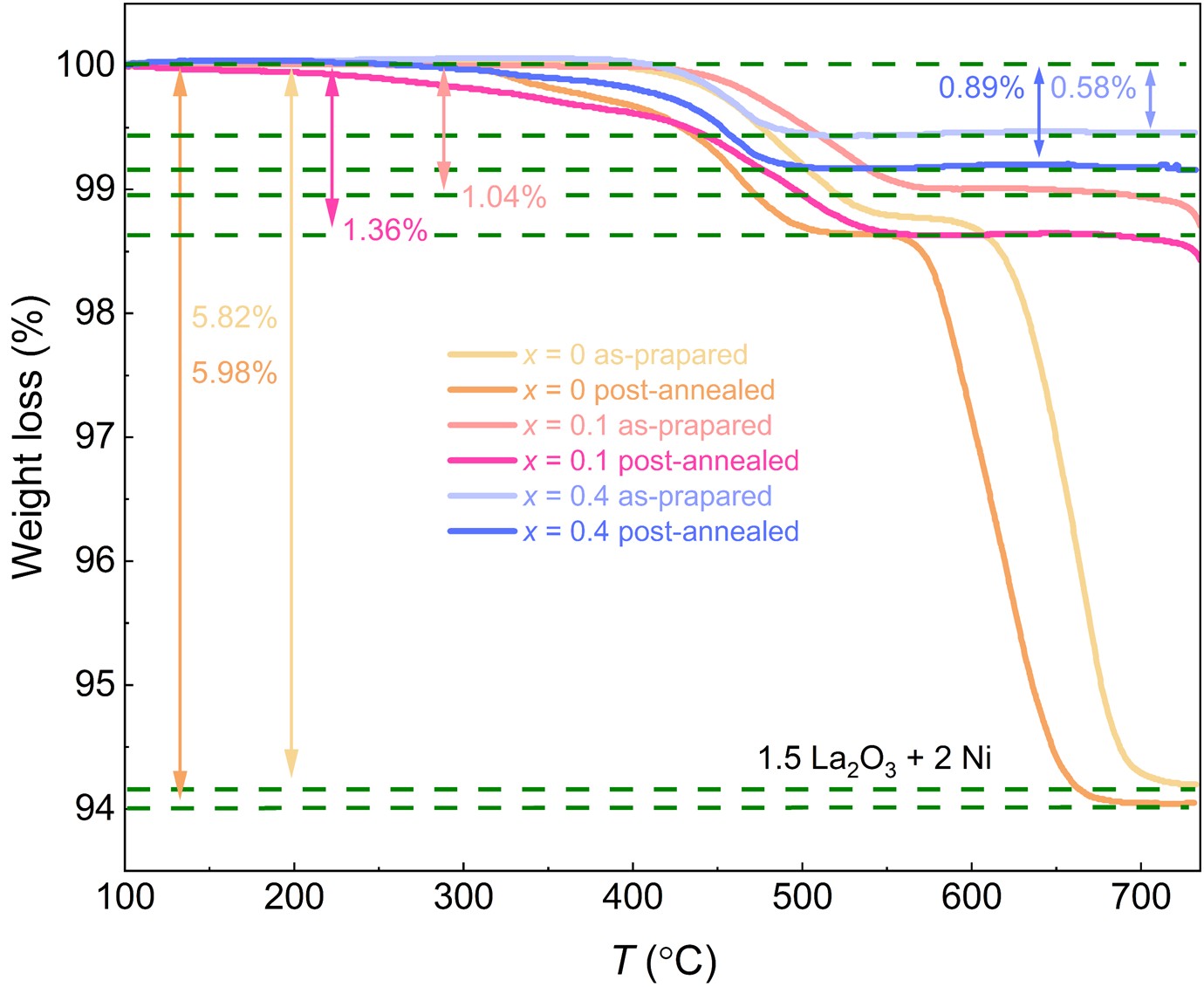}
    \centering
	\caption{Thermogravimetric curves for as-prepared and post-annealed La$_3$Ni$_{2-x}$Al$_x$O$_{7-\delta}$ ($x$ = 0, 0.1, 0.4). The oxygen stoichiometry was determined as La$_3$Ni$_{2}$O$_{6.84}$ for the as-prepared phase and La$_3$Ni$_{2}$O$_{6.91}$ for the post-annealed phase, with the slight deviation from stoichiometry attributed to the presence of about 5~wt.$\%$ La$_2$NiO$_{4}$ in the samples used for TGA. For $x$ = 0.1 samples, analysis of the weight-loss plateaus corresponding to La$_3$Ni$_{1.9}$Al$_{0.1}$O$_{6.45}$ yields oxygen contents of about 6.87 (as-prepared) and 7.00 (post-annealed).}
	\label{figS3}
\end{figure*}

\begin{figure*}[htbp]
	\includegraphics[width=10cm]{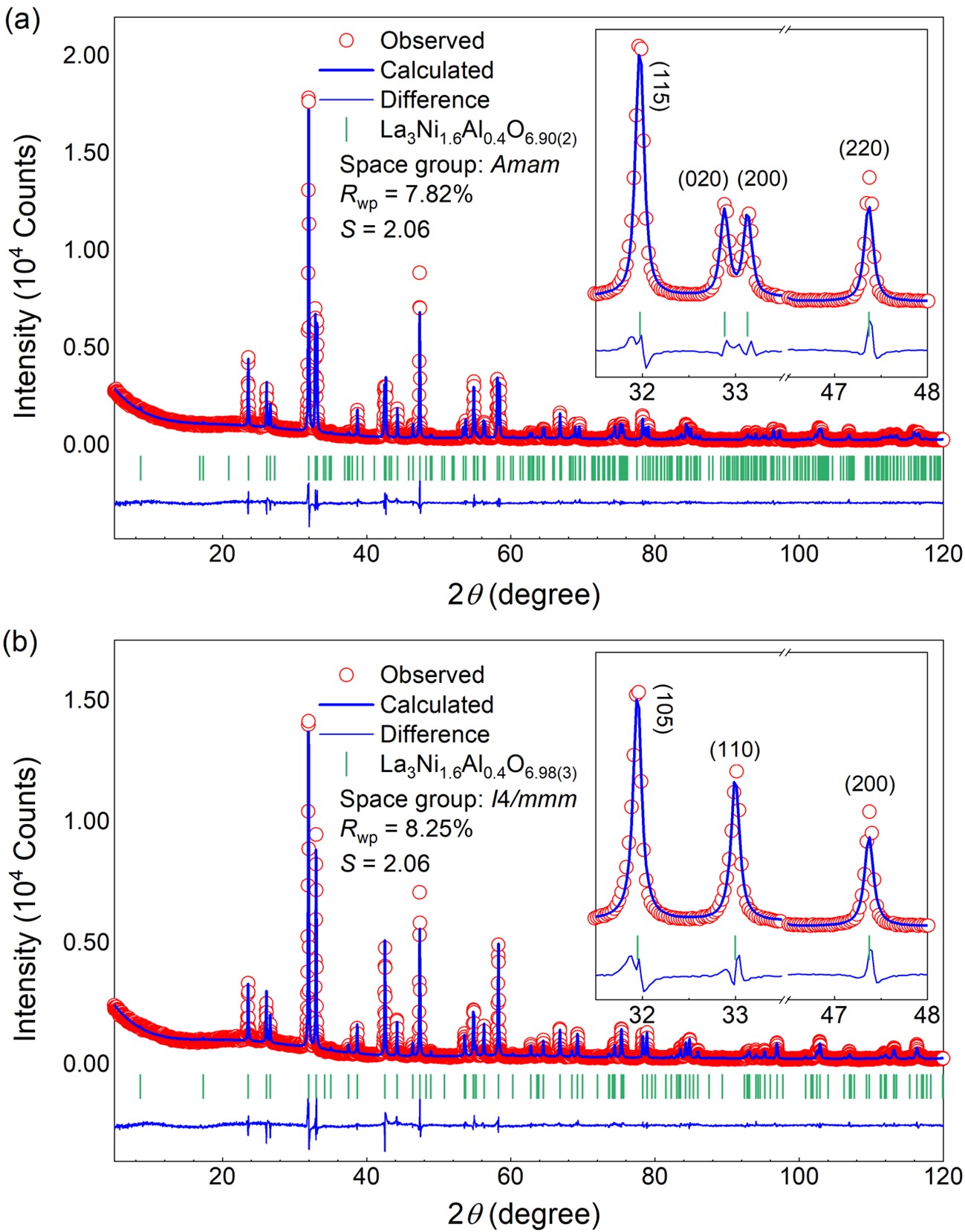}
    \centering
	\caption{Rietveld refinement profiles based on powder X-ray diffraction data for as-prepared La$_3$Ni$_{1.6}$Al$_{0.4}$O$_{6.90(2)}$ (a) and post-annealed La$_3$Ni$_{1.6}$Al$_{0.4}$O$_{6.98(3)}$ (b). The obtained lattice parameters are, $a$ = 5.40466(7)~\r{A}, $b$ = 5.44409(7)~\r{A}, $c$ = 20.4439(3)~\r{A} for the as-prepared La$_3$Ni$_{1.6}$Al$_{0.4}$O$_{6.90(2)}$; and $a$ = 3.83438(4)~\r{A}, $c$ = 20.4620(4)~\r{A} for the post-annealed La$_3$Ni$_{1.6}$Al$_{0.4}$O$_{6.98(3)}$.}
	\label{figS4}
\end{figure*}

\begin{figure*}
	\includegraphics[width=10cm]{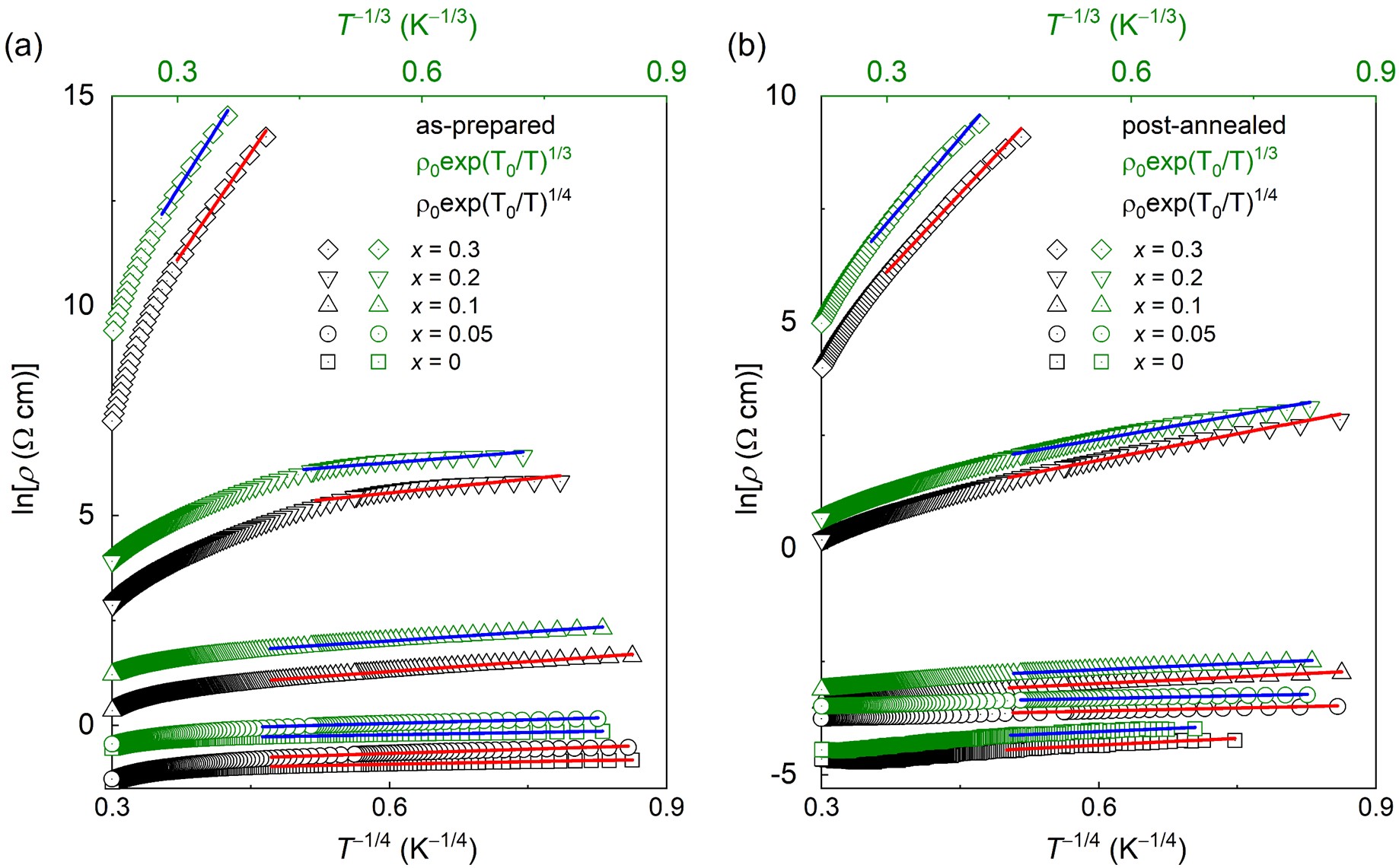}
    \centering
	\caption{The linear fits of ln$\rho$ to $T^{-1/3}$ (blue lines) and $T^{-1/4}$ (red lines) for the as-prepared (a) and post-annealed (b) samples of La$_3$Ni$_{2-x}$Al$_x$O$_{7-\delta}$ ($x \leq$ 0.3) in the low temperature regime. }
	\label{figS5}
\end{figure*}

\begin{figure*}
	\includegraphics[width=10cm]{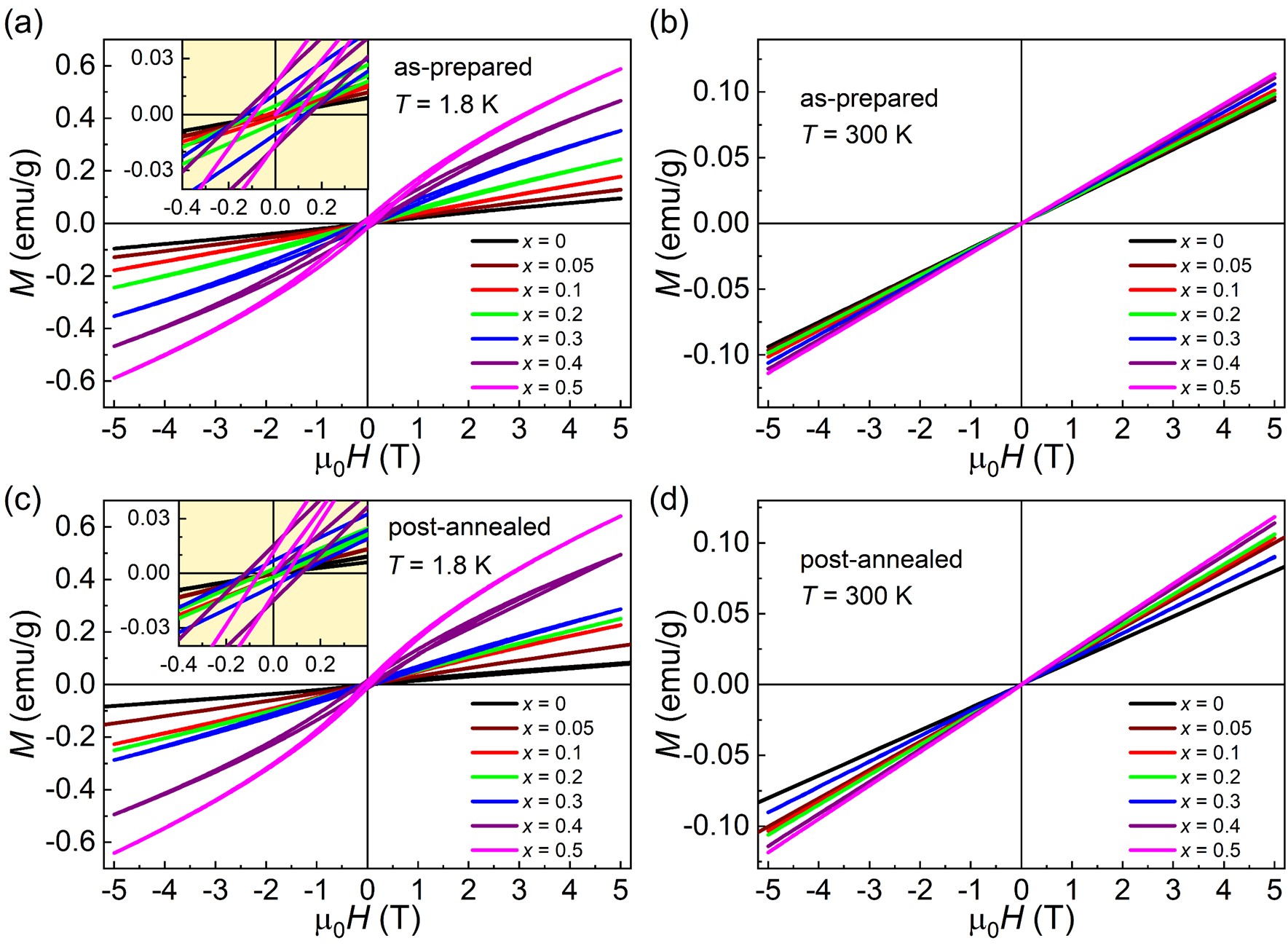}
    \centering
	\caption{Isothermal magnetization for the as-prepared (a, b) and post-annealed (c, d) samples of La$_3$Ni$_{2-x}$Al$_x$O$_{7-\delta}$ at 1.8 K and 300 K.}
	\label{figS6}
\end{figure*}

\begin{figure*}
	\includegraphics[width=8cm]{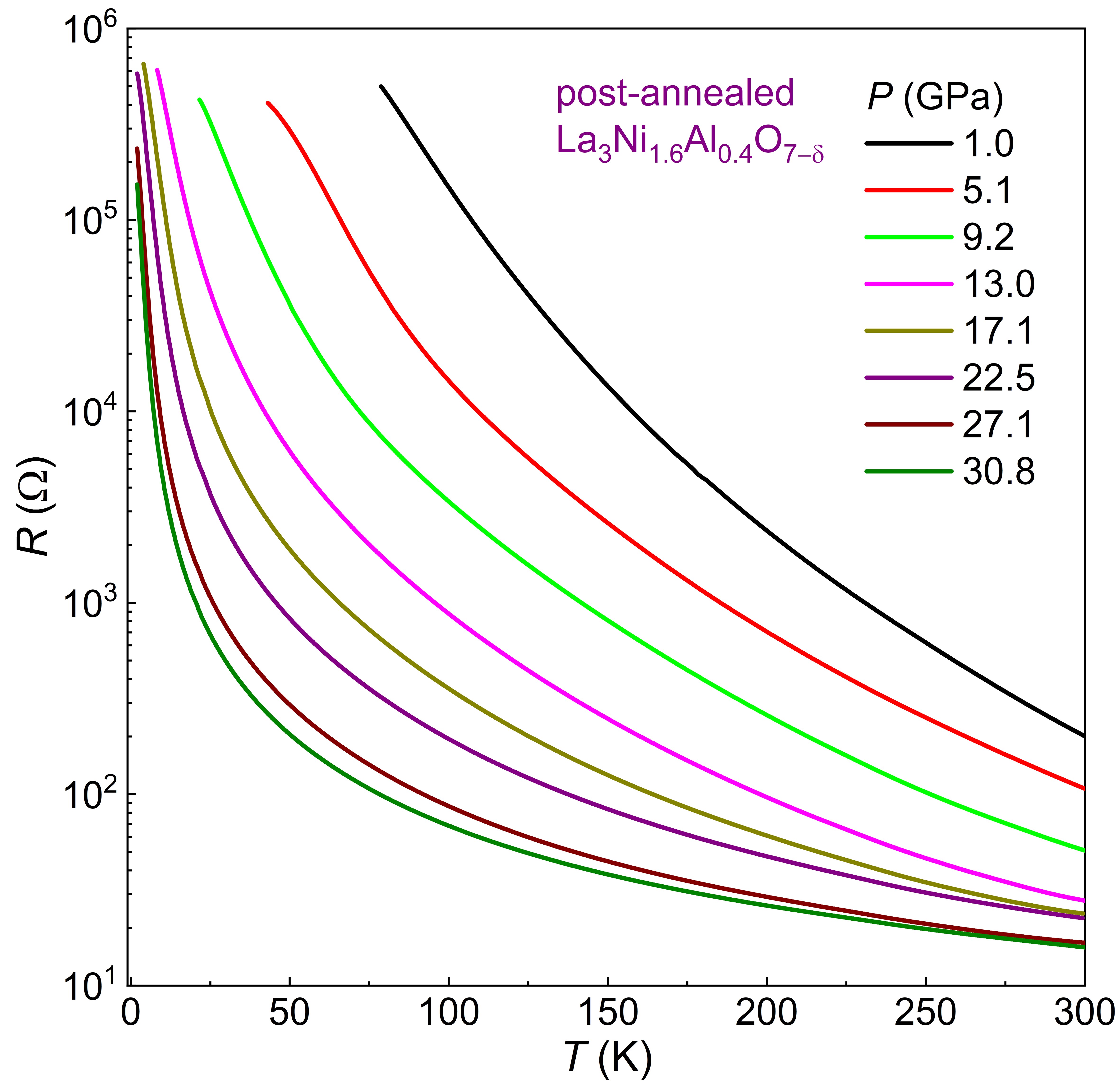}
    \centering
	\caption{Temperature dependence of resistance under pressures for post-annealed La$_3$Ni$_{1.6}$Al$_{0.4}$O$_{7-\delta}$.}
	\label{figS7}
\end{figure*}

\begin{figure*}
	\includegraphics[width=8cm]{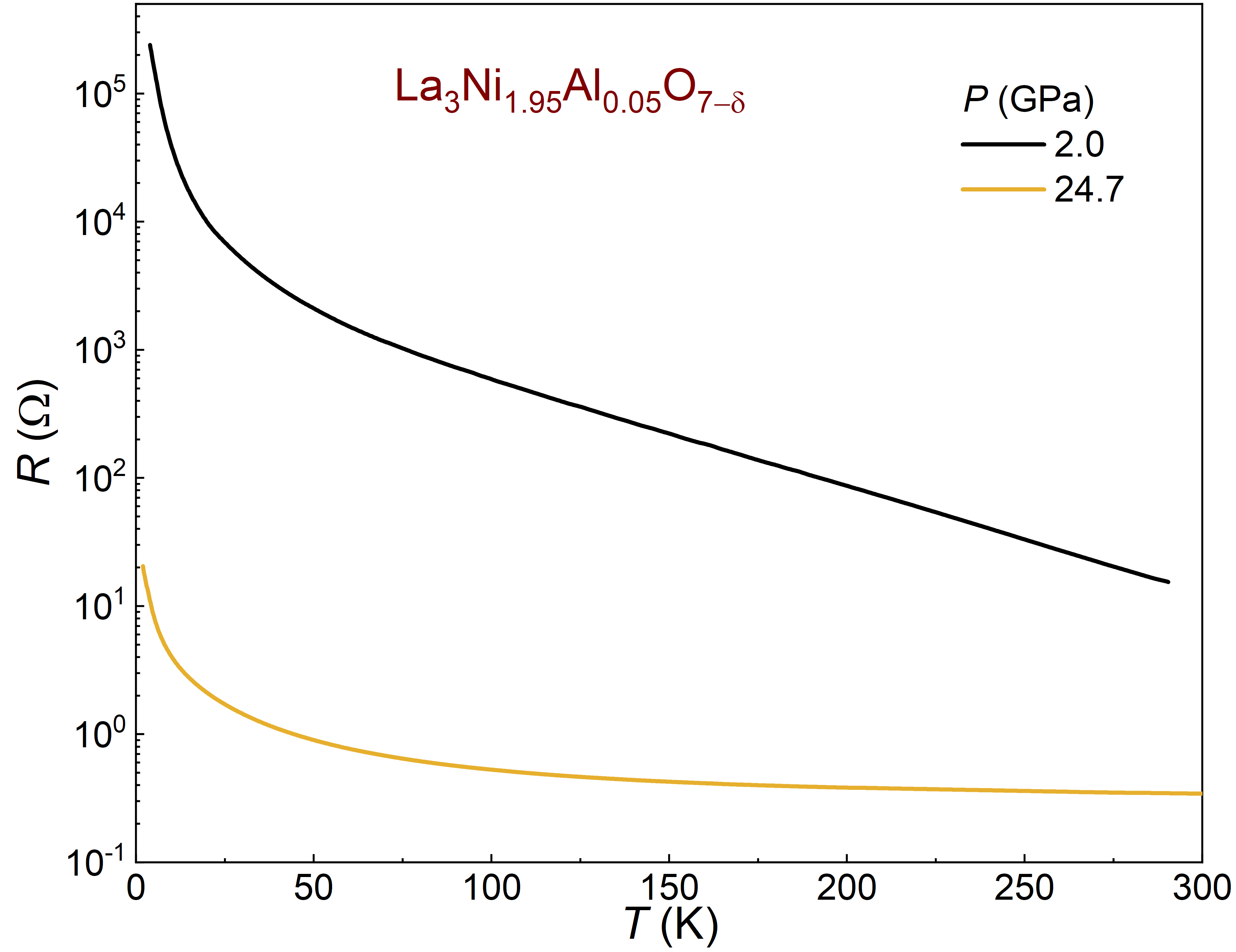}
    \centering
	\caption{Temperature dependence of resistance under pressures for the orthorhombic phase of La$_3$Ni$_{1.95}$Al$_{0.05}$O$_{7-\delta}$ annealed in 10~MPa oxygen at 400~$^\circ$C for 30h.}
	\label{figS8}
\end{figure*}

\clearpage

\begin{table}[htbp]
	\caption{Crystallographic data of the as-prepared ($Amam$ phase) and post-annealed ($I$4$/mmm$ phase) La$_3$Ni$_{1.6}$Al$_{0.4}$O$_{7-\delta}$ from X-ray powder diffractions.}
	\label{tableS1}
	\begin{tabular}{cccccc}
		\toprule
		\multicolumn{3}{c}{La$_3$Ni$_{1.6}$Al$_{0.4}$O$_{6.90}$}  &\multicolumn{3}{c}{$Amam$ (No.63)}   \\
		\midrule
		atom  & $x$     & $y$       & $z$        &  occ.  &  $U_{iso}$                    \\
		% \hline
		La(1)    & 0.25     & 0.249(2)    & 0.5           & 1.00        & 0.004(1)               \\
		La(2)    & 0.25     & 0.242(1)    & 0.3195(1)     & 1.00        & 0.004(1)                     \\
		Ni/Al    & 0.25     & 0.248(4)    & 0.0971(3)   & 0.80/0.20   & 0.007(2)                \\
		O(1)     & 0.25     & 0.288(14)    & 0             & 0.90(2)     & 0.013(13)                \\
		O(2)     & 0.25     & 0.311(5)    & 0.1996(10)   & 1.00        & 0.015(8)              \\
		O(3)     & 0.5      & 0           & 0.105(2)   & 1.00        & 0.001(11)               \\
		O(4)     & 0        & 0.5         & 0.091(2)   & 1.00        & 0.015(8)                     \\
		\midrule\midrule
		\multicolumn{3}{c}{La$_3$Ni$_{1.6}$Al$_{0.4}$O$_{6.98}$}  &\multicolumn{3}{c}{$I$4$/mmm$ (No.139)}   \\
		\midrule
		atom       & $x$     & $y$       & $z$        &  occ.  &  $U_{iso}$                    \\
		% \hline
		La(1)      & 0     & 0       & 0.3198(2)     & 1.00     & 0.004(1)             \\
		La(2)      & 0     & 0       & 0.5           & 1.00     & 0.004(1)                    \\
		Ni/Al      & 0     & 0       & 0.0977(4)     & 0.80/0.20     & 0.016(2)                 \\
		O(1)       & 0     & 0       & 0             & 0.98(3)  & 0.012(11)                \\
		O(2)       & 0     & 0       & 0.191(2)     & 1.00     & 0.051(13)               \\
		O(3)       & 0     & 0.5     & 0.0965(10)   & 1.00     & 0.033(7)               \\
		\bottomrule
	\end{tabular}
\end{table}

\begin{table}[htbp]
	\caption{The Curie-Weiss fit parameters for La$_3$Ni$_{2-x}$Al$_x$O$_{7-\delta}$ (0.2 $\leq x\leq$ 0.5).}
	\label{tableS1}
	\begin{tabular}{ccccccc}
		\toprule
		\multicolumn{3}{c}{doping content (as-prepared)}  &0.2  &0.3  &0.4   &0.5    \\
        \multicolumn{3}{c}{$\chi_0$ (emu mol$^{-1}$)}  &0.00011  &0.0011  &0.0010   &0.0010    \\
        \multicolumn{3}{c}{$\theta_\mathrm{W}$ (K)}  &8.50  &30.96  &17.78   &27.02    \\			
        \multicolumn{3}{c}{$C$ (emu K mol$^{-1}$ fu$^{-1}$)}  &0.051	&0.12	&0.14	&0.19    \\
        \multicolumn{3}{c}{$\mu_\mathrm{eff}$ ($\mu_{\rm B}$/Ni)}  &0.48	&0.74	&0.84	&1.01    \\
		\midrule\midrule
        \multicolumn{3}{c}{doping content (post-annealed)}  &0.2  &0.3  &0.4   &0.5    \\
        \multicolumn{3}{c}{$\chi_0$ (emu mol$^{-1}$)}  &0.00096	&0.00095	&0.00095	&0.00092    \\
        \multicolumn{3}{c}{$\theta_\mathrm{W}$ (K)}  &100	&84.68	&50.72	&49.98    \\
        \multicolumn{3}{c}{$C$ (emu K mol$^{-1}$ fu$^{-1}$)}  &0.12	&0.18	&0.19	&0.24    \\
        \multicolumn{3}{c}{$\mu_\mathrm{eff}$ ($\mu_{\rm B}$/Ni)}  &0.74	&0.92	&0.98	&1.14   \\
		\bottomrule
	\end{tabular}
\end{table}

\end{appendices}

\clearpage

%\bibliography{references}

\begin{thebibliography}{63}
\ifx \bisbn   \undefined \def \bisbn  #1{ISBN #1}\fi
\ifx \binits  \undefined \def \binits#1{#1}\fi
\ifx \bauthor  \undefined \def \bauthor#1{#1}\fi
\ifx \batitle  \undefined \def \batitle#1{#1}\fi
\ifx \bjtitle  \undefined \def \bjtitle#1{#1}\fi
\ifx \bvolume  \undefined \def \bvolume#1{\textbf{#1}}\fi
\ifx \byear  \undefined \def \byear#1{#1}\fi
\ifx \bissue  \undefined \def \bissue#1{#1}\fi
\ifx \bfpage  \undefined \def \bfpage#1{#1}\fi
\ifx \blpage  \undefined \def \blpage #1{#1}\fi
\ifx \burl  \undefined \def \burl#1{\textsf{#1}}\fi
\ifx \doiurl  \undefined \def \doiurl#1{\url{https://doi.org/#1}}\fi
\ifx \betal  \undefined \def \betal{\textit{et al.}}\fi
\ifx \binstitute  \undefined \def \binstitute#1{#1}\fi
\ifx \binstitutionaled  \undefined \def \binstitutionaled#1{#1}\fi
\ifx \bctitle  \undefined \def \bctitle#1{#1}\fi
\ifx \beditor  \undefined \def \beditor#1{#1}\fi
\ifx \bpublisher  \undefined \def \bpublisher#1{#1}\fi
\ifx \bbtitle  \undefined \def \bbtitle#1{#1}\fi
\ifx \bedition  \undefined \def \bedition#1{#1}\fi
\ifx \bseriesno  \undefined \def \bseriesno#1{#1}\fi
\ifx \blocation  \undefined \def \blocation#1{#1}\fi
\ifx \bsertitle  \undefined \def \bsertitle#1{#1}\fi
\ifx \bsnm \undefined \def \bsnm#1{#1}\fi
\ifx \bsuffix \undefined \def \bsuffix#1{#1}\fi
\ifx \bparticle \undefined \def \bparticle#1{#1}\fi
\ifx \barticle \undefined \def \barticle#1{#1}\fi
\bibcommenthead
\ifx \bconfdate \undefined \def \bconfdate #1{#1}\fi
\ifx \botherref \undefined \def \botherref #1{#1}\fi
\ifx \url \undefined \def \url#1{\textsf{#1}}\fi
\ifx \bchapter \undefined \def \bchapter#1{#1}\fi
\ifx \bbook \undefined \def \bbook#1{#1}\fi
\ifx \bcomment \undefined \def \bcomment#1{#1}\fi
\ifx \oauthor \undefined \def \oauthor#1{#1}\fi
\ifx \citeauthoryear \undefined \def \citeauthoryear#1{#1}\fi
\ifx \endbibitem  \undefined \def \endbibitem {}\fi
\ifx \bconflocation  \undefined \def \bconflocation#1{#1}\fi
\ifx \arxivurl  \undefined \def \arxivurl#1{\textsf{#1}}\fi
\csname PreBibitemsHook\endcsname

%%% 1
\bibitem[\protect\citeauthoryear{Sun et~al.}{2023}]{Sun2023}
\begin{barticle}
\bauthor{\bsnm{Sun}, \binits{H.}}, \betal:
\batitle{Signatures of superconductivity near 80~{K} in a nickelate under high
  pressure}.
\bjtitle{$Nature$}
\bvolume{621}(\bissue{621}),
\bfpage{493}--\blpage{498}
(\byear{2023})
\end{barticle}
\endbibitem

%%% 2
\bibitem[\protect\citeauthoryear{Wang et~al.}{2024}]{2024PRX.ChengJG}
\begin{barticle}
\bauthor{\bsnm{Wang}, \binits{G.}}, \betal:
\batitle{Pressure-induced superconductivity in polycrystalline
  {La$_3$Ni$_2$O$_{7-\delta}$}}.
\bjtitle{$Phys.$ $Rev.$ $X$}
\bvolume{14},
\bfpage{011040}
(\byear{2024})
\end{barticle}
\endbibitem

%%% 3
\bibitem[\protect\citeauthoryear{Zhang et~al.}{2024}]{2024NP.YuanHQ}
\begin{barticle}
\bauthor{\bsnm{Zhang}, \binits{Y.}}, \betal:
\batitle{High-temperature superconductivity with zero resistance and
  strange-metal behaviour in {La$_3$Ni$_2$O$_{7-\delta}$}}.
\bjtitle{$Nat.$ $Phys.$}
\bvolume{20}(\bissue{8}),
\bfpage{1269}--\blpage{1273}
(\byear{2024})
\end{barticle}
\endbibitem

%%% 4
\bibitem[\protect\citeauthoryear{Li et~al.}{2019}]{Li2019}
\begin{barticle}
\bauthor{\bsnm{Li}, \binits{D.}}, \betal:
\batitle{Superconductivity in an infinite-layer nickelate}.
\bjtitle{$Nature$}
\bvolume{572}(\bissue{7771}),
\bfpage{624}--\blpage{627}
(\byear{2019})
\end{barticle}
\endbibitem

%%% 5
\bibitem[\protect\citeauthoryear{Shen et~al.}{2023}]{2023CPL.ZhangGM}
\begin{botherref}
\oauthor{\bsnm{Shen}, \binits{Y.}},
\oauthor{\bsnm{Qin}, \binits{M.}},
\oauthor{\bsnm{Zhang}, \binits{G.-M.}}:
Effective bi-layer model hamiltonian and density-matrix renormalization group
  study for the {High-$T_\mathrm{c}$} superconductivity in {La$_3$Ni$_2$O$_7$}
  under high pressure.
$Chin.$ $Phys.$ $Lett.$
\textbf{40}(12)
(2023)
\end{botherref}
\endbibitem

%%% 6
\bibitem[\protect\citeauthoryear{Lechermann et~al.}{2023}]{Lechermann2023}
\begin{barticle}
\bauthor{\bsnm{Lechermann}, \binits{F.}}, \betal:
\batitle{Electronic correlations and superconducting instability in
  {La$_3$Ni$_2$O$_7$} under high pressure}.
\bjtitle{$Phys.$ $Rev.$ $B$}
\bvolume{108},
\bfpage{201121}
(\byear{2023})
\end{barticle}
\endbibitem

%%% 7
\bibitem[\protect\citeauthoryear{Zhang et~al.}{2023}]{2023PRB.Dagotto}
\begin{barticle}
\bauthor{\bsnm{Zhang}, \binits{Y.}}, \betal:
\batitle{Electronic structure, dimer physics, orbital-selective behavior, and
  magnetic tendencies in the bilayer nickelate superconductor
  {La$_3$Ni$_2$O$_7$} under pressure}.
\bjtitle{$Phys.$ $Rev.$ $B$}
\bvolume{108},
\bfpage{180510}
(\byear{2023})
\end{barticle}
\endbibitem

%%% 8
\bibitem[\protect\citeauthoryear{Yang et~al.}{2023}]{2023PRB.YangYF}
\begin{barticle}
\bauthor{\bsnm{Yang}, \binits{Y.}},
\bauthor{\bsnm{Zhang}, \binits{G.}},
\bauthor{\bsnm{Zhang}, \binits{F.}}:
\batitle{Interlayer valence bonds and two-component theory for
  {high-$T_\mathrm{c}$} superconductivity of {La$_3$Ni$_2$O$_7$} under
  pressure}.
\bjtitle{$Phys.$ $Rev.$ $B$}
\bvolume{108},
\bfpage{201108}
(\byear{2023})
\end{barticle}
\endbibitem

%%% 9
\bibitem[\protect\citeauthoryear{Liu et~al.}{2023}]{YangF2023.PRL}
\begin{barticle}
\bauthor{\bsnm{Liu}, \binits{Y.}}, \betal:
\batitle{${s}^{\ifmmode\pm\else\textpm\fi{}}$-wave pairing and the destructive
  role of apical-oxygen deficiencies in {La$_3$Ni$_2$O$_7$} under pressure}.
\bjtitle{$Phys.$ $Rev.$ $Lett.$}
\bvolume{131},
\bfpage{236002}
(\byear{2023})
\end{barticle}
\endbibitem

%%% 10
\bibitem[\protect\citeauthoryear{Sakakibara et~al.}{2024}]{2024PRL.Kuroki}
\begin{barticle}
\bauthor{\bsnm{Sakakibara}, \binits{H.}},
\bauthor{\bsnm{Kitamine}, \binits{N.}},
\bauthor{\bsnm{Ochi}, \binits{M.}},
\bauthor{\bsnm{Kuroki}, \binits{K.}}:
\batitle{Possible high {$T_\mathrm{c}$} superconductivity in
  {La$_3$Ni$_2$O$_7$} under high pressure through manifestation of a nearly
  half-filled bilayer hubbard model}.
\bjtitle{$Phys.$ $Rev.$ $Lett.$}
\bvolume{132},
\bfpage{106002}
(\byear{2024})
\end{barticle}
\endbibitem

%%% 11
\bibitem[\protect\citeauthoryear{Geisler et~al.}{2024a}]{Geisler2024}
\begin{barticle}
\bauthor{\bsnm{Geisler}, \binits{B.}},
\bauthor{\bsnm{Hamlin}, \binits{J.J.}},
\bauthor{\bsnm{Stewart}, \binits{G.R.}},
\bauthor{\bsnm{Hennig}, \binits{R.G.}},
\bauthor{\bsnm{Hirschfeld}, \binits{P.J.}}:
\batitle{Structural transitions, octahedral rotations, and electronic
  properties of {$A_3$Ni$_2$O$_7$} rare-earth nickelates under high pressure}.
\bjtitle{$npj$ $Quantum$ $Mater.$}
\bvolume{9}(\bissue{9}),
\bfpage{38}
(\byear{2024})
\end{barticle}
\endbibitem

%%% 12
\bibitem[\protect\citeauthoryear{Geisler et~al.}{2024b}]{2024npj.Geisler-2}
\begin{barticle}
\bauthor{\bsnm{Geisler}, \binits{B.}}, \betal:
\batitle{Optical properties and electronic correlations in {La$_3$Ni$_2$O$_7$}
  bilayer nickelates under high pressure}.
\bjtitle{$npj$ $Quantum$ $Mater.$}
\bvolume{9}(\bissue{9}),
\bfpage{89}
(\byear{2024})
\end{barticle}
\endbibitem

%%% 13
\bibitem[\protect\citeauthoryear{Lu et~al.}{2024}]{2024WuCJ.PRL}
\begin{barticle}
\bauthor{\bsnm{Lu}, \binits{C.}},
\bauthor{\bsnm{Pan}, \binits{Z.}},
\bauthor{\bsnm{Yang}, \binits{F.}},
\bauthor{\bsnm{Wu}, \binits{C.}}:
\batitle{Interlayer-coupling-driven high-temperature superconductivity in
  {${\mathrm{La}}_{3}{\mathrm{Ni}}_{2}{\mathrm{O}}_{7}$} under pressure}.
\bjtitle{$Phys.$ $Rev.$ $Lett.$}
\bvolume{132},
\bfpage{146002}
(\byear{2024})
\end{barticle}
\endbibitem

%%% 14
\bibitem[\protect\citeauthoryear{Qu et~al.}{2024}]{SuG2024.PRL}
\begin{barticle}
\bauthor{\bsnm{Qu}, \binits{X.}}, \betal:
\batitle{Bilayer
  ${t\text{\ensuremath{-}}J\text{\ensuremath{-}}J}_{\ensuremath{\perp}}$ model
  and magnetically mediated pairing in the pressurized nickelate
  {${\mathrm{La}}_{3}{\mathrm{Ni}}_{2}{\mathrm{O}}_{7}$}}.
\bjtitle{$Phys.$ $Rev.$ $Lett.$}
\bvolume{132},
\bfpage{036502}
(\byear{2024})
\end{barticle}
\endbibitem

%%% 15
\bibitem[\protect\citeauthoryear{Deng et~al.}{2025}]{2024PNAS.Deng}
\begin{barticle}
\bauthor{\bsnm{Deng}, \binits{L.}}, \betal:
\batitle{Creation, stabilization, and investigation at ambient pressure of
  pressure-induced superconductivity in {Bi$_{0.5}$Sb$_{1.5}$Te$_3$}}.
\bjtitle{$PNAS$}
\bvolume{122}(\bissue{6}),
\bfpage{2423102122}
(\byear{2025})
\end{barticle}
\endbibitem

%%% 16
\bibitem[\protect\citeauthoryear{Xu et~al.}{2024}]{2024AEM.Xu}
\begin{barticle}
\bauthor{\bsnm{Xu}, \binits{M.}}, \betal:
\batitle{Pressure-dependent ``insulator--metal--insulator'' behavior in Sr-doped La$_3$Ni$_2$O$_7$}.
\bjtitle{Advanced Electronic Materials}
\bvolume{10}(\bissue{9}),
\bfpage{2400078}
(\byear{2024})
\end{barticle}
\endbibitem

%%% 17
\bibitem[\protect\citeauthoryear{Drennan et~al.}{1982}]{Drennan1982}
\begin{barticle}
\bauthor{\bsnm{Drennan}, \binits{J.}},
\bauthor{\bsnm{Tavares}, \binits{C.P.}},
\bauthor{\bsnm{Steele}, \binits{B.C.H.}}:
\batitle{An electron microscope investigation of phases in the system
  {La-Ni-O}}.
\bjtitle{$Mater.$ $Res.$ $Bull.$}
\bvolume{17}(\bissue{5}),
\bfpage{621}--\blpage{626}
(\byear{1982})
\end{barticle}
\endbibitem

%%% 18
\bibitem[\protect\citeauthoryear{Zhang et~al.}{1994}]{Zhang1994}
\begin{barticle}
\bauthor{\bsnm{Zhang}, \binits{Z.}},
\bauthor{\bsnm{Greenblatt}, \binits{M.}},
\bauthor{\bsnm{Goodenough}, \binits{J.}}:
\batitle{Synthesis, structure, and properties of the layered perovskite
  {La$_3$Ni$_2$O$_{7-\delta}$}}.
\bjtitle{$J.$ $Solid$ $State$ $Chem.$}
\bvolume{108}(\bissue{2}),
\bfpage{402}--\blpage{409}
(\byear{1994})
\end{barticle}
\endbibitem

%%% 19
\bibitem[\protect\citeauthoryear{Sreedhar et~al.}{1994}]{SREEDHAR1994}
\begin{barticle}
\bauthor{\bsnm{Sreedhar}, \binits{K.}}, \betal:
\batitle{Low-temperature electronic properties of the
  {La$_{n+1}$Ni$_n$O$_{3n+1}$} ($n$ = 2, 3, and $\infty$) system: Evidence for
  a crossover from fluctuating-valence to fermi-liquid-like behavior}.
\bjtitle{$J.$ $Solid$ $State$ $Chem.$}
\bvolume{110}(\bissue{2}),
\bfpage{208}--\blpage{215}
(\byear{1994})
\end{barticle}
\endbibitem

%%% 20
\bibitem[\protect\citeauthoryear{Taniguchi et~al.}{1995}]{Taniguchi1995}
\begin{barticle}
\bauthor{\bsnm{Taniguchi}, \binits{S.}}, \betal:
\batitle{Transport, magnetic and thermal properties of
  {La$_3$Ni$_2$O$_{7-\delta}$}}.
\bjtitle{$J.$ $Phys.$ $Soc.$ $Jpn.$}
\bvolume{64}(\bissue{5}),
\bfpage{1644}--\blpage{1650}
(\byear{1995})
\end{barticle}
\endbibitem

%%% 21
\bibitem[\protect\citeauthoryear{Ling et~al.}{2000}]{Ling2000}
\begin{barticle}
\bauthor{\bsnm{Ling}, \binits{C.}},
\bauthor{\bsnm{Argyriou}, \binits{D.}},
\bauthor{\bsnm{Wu}, \binits{G.}},
\bauthor{\bsnm{J.}, \binits{N.}}:
\batitle{Neutron diffraction study of {La$_3$Ni$_2$O$_7$}: Structural
  relationships among $n$ = 1, 2, and 3 phases {La$_{n+1}$Ni$_n$O$_{3n+1}$}}.
\bjtitle{$J.$ $Solid$ $State$ $Chem.$}
\bvolume{152}(\bissue{2}),
\bfpage{517}--\blpage{525}
(\byear{2000})
\end{barticle}
\endbibitem

%%% 22
\bibitem[\protect\citeauthoryear{Wang et~al.}{2024}]{Structure2024}
\begin{barticle}
\bauthor{\bsnm{Wang}, \binits{L.}}, \betal:
\batitle{Structure responsible for the superconducting state in
  {La$_3$Ni$_2$O$_7$} at high-pressure and low-temperature conditions}.
\bjtitle{$J.$ $Am.$ $Chem.$ $Soc.$}
\bvolume{146}(\bissue{11}),
\bfpage{7506}--\blpage{7514}
(\byear{2024})
\end{barticle}
\endbibitem

%%% 23
\bibitem[\protect\citeauthoryear{Li et~al.}{2025}]{327.100GPa}
\begin{botherref}
\oauthor{\bsnm{Li}, \binits{J.}}, et al.:
Identification of superconductivity in bilayer nickelate {La$_3$Ni$_2$O$_7$}
  under high pressure up to 100~{GPa}.
$Natl.$ $Sci.$ $Rev.$,
220
(2025)
\end{botherref}
\endbibitem

%%% 24
\bibitem[\protect\citeauthoryear{Wang et~al.}{2024}]{Wang2024}
\begin{barticle}
\bauthor{\bsnm{Wang}, \binits{N.}}, \betal:
\batitle{Bulk high-temperature superconductivity in pressurized tetragonal
  {La$_2$PrNi$_2$O$_7$}}.
\bjtitle{$Nature$}
\bvolume{634},
\bfpage{579}--\blpage{584}
(\byear{2024})
\end{barticle}
\endbibitem

%%% 25
\bibitem[\protect\citeauthoryear{Zhu et~al.}{2024}]{Zhu2024}
\begin{barticle}
\bauthor{\bsnm{Zhu}, \binits{Y.}}, \betal:
\batitle{Superconductivity in pressurized trilayer
  {La$_4$Ni$_3$O$_{10-\delta}$} single crystals}.
\bjtitle{$Nature$}
\bvolume{631},
\bfpage{531}--\blpage{536}
(\byear{2024})
\end{barticle}
\endbibitem

%%% 26
\bibitem[\protect\citeauthoryear{Ko et~al.}{2025}]{APSC.Hwang-1}
\begin{barticle}
\bauthor{\bsnm{Ko}, \binits{E.}}, \betal:
\batitle{Signatures of ambient pressure superconductivity in thin film
  {La$_3$Ni$_2$O$_7$}}.
\bjtitle{$Nature$}
\bvolume{638},
\bfpage{935}--\blpage{940}
(\byear{2025})
\end{barticle}
\endbibitem

%%% 27
\bibitem[\protect\citeauthoryear{Zhou et~al.}{2025}]{Zhou2025.nature}
\begin{barticle}
\bauthor{\bsnm{Zhou}, \binits{G.}}, \betal:
\batitle{Ambient-pressure superconductivity onset above 40 {K} in
  {(La,Pr)$_3$Ni$_2$O$_7$} films}.
\bjtitle{$Nature$}
\bvolume{640},
\bfpage{641}--\blpage{646}
(\byear{2025})
\end{barticle}
\endbibitem

%%% 28
\bibitem[\protect\citeauthoryear{Liu et~al.}{2025}]{APSC.Hwang-2}
\begin{barticle}
\bauthor{\bsnm{Liu}, \binits{Y.}}, \betal:
\batitle{Superconductivity and normal-state transport in compressively strained
  {La$_2$PrNi$_2$O$_7$} thin films}.
\bjtitle{$Nat.$ $Mater.$}
\bvolume{24},
\bfpage{1221}--\blpage{1227}
(\byear{2025})
\end{barticle}
\endbibitem

%%% 29
\bibitem[\protect\citeauthoryear{Goldschmidt}{1926}]{Goldschmidt1926}
\begin{barticle}
\bauthor{\bsnm{Goldschmidt}, \binits{V.M.}}:
\batitle{Die gesetze der krystallochemie}.
\bjtitle{$Naturwissenschaften$}
\bvolume{14},
\bfpage{477}--\blpage{485}
(\byear{1926})
\end{barticle}
\endbibitem

%%% 30
\bibitem[\protect\citeauthoryear{Li et~al.}{2016}]{2016CM.tolerance}
\begin{barticle}
\bauthor{\bsnm{Li}, \binits{Z.}}, \betal:
\batitle{Stabilizing perovskite structures by tuning tolerance factor:
  Formation of formamidinium and cesium lead iodide solid-state alloys}.
\bjtitle{$Chem.$ $Mater.$}
\bvolume{28}(\bissue{1}),
\bfpage{284}--\blpage{292}
(\byear{2016})
\end{barticle}
\endbibitem

%%% 31
\bibitem[\protect\citeauthoryear{Shannon}{1976}]{Shannon1976}
\begin{barticle}
\bauthor{\bsnm{Shannon}, \binits{R.D.}}:
\batitle{Revised effective ionic-radii and systematic studies of interatomic
  distances in halides and chalcogenides}.
\bjtitle{$Acta$ $Crystallogr.,$ $Sect.$ $A$}
\bvolume{32},
\bfpage{751}--\blpage{767}
(\byear{1976})
\end{barticle}
\endbibitem

%%% 32
\bibitem[\protect\citeauthoryear{Kharlamova et~al.}{2018}]{Inga2018}
\begin{barticle}
\bauthor{\bsnm{Kharlamova}, \binits{I.}}, \betal:
\batitle{Ruddlesden-popper phases {Sr$_3$Ni$_{2-x}$Al$_x$O$_{7-\delta}$} and
  some doped derivatives: Synthesis, oxygen nonstoichiometry and electrical
  properties}.
\bjtitle{$Solid$ $State$ $Ion.$}
\bvolume{324},
\bfpage{241}--\blpage{246}
(\byear{2018})
\end{barticle}
\endbibitem

%%% 33
\bibitem[\protect\citeauthoryear{Yilmaz et~al.}{2024}]{Yilmaz2024}
\begin{barticle}
\bauthor{\bsnm{Yilmaz}, \binits{H.}}, \betal:
\batitle{Realization of a classical ruddlesden popper type bilayer nickelate in
  {Sr$_3$Ni$_{2-x}$Al$_x$O$_{7-\delta}$ with unusual Ni$^{4+}$}}.
\bjtitle{$npj$ $Quantum$ $Mater.$}
\bvolume{9},
\bfpage{92}--\blpage{100}
(\byear{2024})
\end{barticle}
\endbibitem

%%% 34
\bibitem[\protect\citeauthoryear{Yamane et~al.}{2025}]{yamane2024}
\begin{barticle}
\bauthor{\bsnm{Yamane}, \binits{K.}}, \betal:
\batitle{High-pressure synthesis of bilayer nickelate
  {Sr$_{3}$Ni$_{2}$O$_{5}$Cl$_{2}$} with a tetragonal crystal structure}.
\bjtitle{$Acta$ $Cryst.$ $C$}
\bvolume{81},
\bfpage{259}--\blpage{263}
(\byear{2025})
\end{barticle}
\endbibitem

%%% 35
\bibitem[\protect\citeauthoryear{Rhodes and Wahl}{2024}]{Rhodes2024}
\begin{barticle}
\bauthor{\bsnm{Rhodes}, \binits{L.C.}},
\bauthor{\bsnm{Wahl}, \binits{P.}}:
\batitle{Structural routes to stabilize superconducting {La$_3$Ni$_2$O$_7$} at
  ambient pressure}.
\bjtitle{$Phys.$ $Rev.$ $Mater.$}
\bvolume{8},
\bfpage{044801}
(\byear{2024})
\end{barticle}
\endbibitem

%%% 36
\bibitem[\protect\citeauthoryear{Wu et~al.}{2024}]{wu2024}
\begin{botherref}
\oauthor{\bsnm{Wu}, \binits{S.}}, et al.:
Ac$_3$Ni$_2$O$_7$ and La$_2$$Ae$Ni$_2$O$_6$F ($Ae$ = Sr, Ba): Benchmark
  Materials for Bilayer Nickelate Superconductivity
(2024).
\url{https://arxiv.org/abs/2403.11713}
\end{botherref}
\endbibitem

%%% 37
\bibitem[\protect\citeauthoryear{Zhang and Greenblatt}{1994}]{Zhang1994-2}
\begin{barticle}
\bauthor{\bsnm{Zhang}, \binits{Z.}},
\bauthor{\bsnm{Greenblatt}, \binits{M.}}:
\batitle{Synthesis, structure, and physical-properties of
  {La$_{3-x}M_x$Ni$_2$O$_{7-\delta}$ ($M=$ Ca$^{2+}$, Sr$^{2+}$, Ba$^{2+}$, $0
  \leq x \leq 0.075$)}}.
\bjtitle{$J.$ $Solid$ $State$ $Chem.$}
\bvolume{111}(\bissue{1}),
\bfpage{141}--\blpage{146}
(\byear{1994})
\end{barticle}
\endbibitem

%%% 38
\bibitem[\protect\citeauthoryear{Aggarwal and
  Bo\v{z}ovi\'{c}}{2024}]{2024Bozovic}
\begin{botherref}
\oauthor{\bsnm{Aggarwal}, \binits{L.}},
\oauthor{\bsnm{Bo\v{z}ovi\'{c}}, \binits{I.}}:
The quest for high-temperature superconductivity in nickelates under ambient
  pressure.
$Materials$
\textbf{17}(11)
(2024)
\end{botherref}
\endbibitem

%%% 39
\bibitem[\protect\citeauthoryear{Periyasamy et~al.}{2021}]{periyasamy2021}
\begin{barticle}
\bauthor{\bsnm{Periyasamy}, \binits{M.}}, \betal:
\batitle{Effect of electron doping on the crystal structure and physical
  properties of an n = 3 {Ruddlesden--Popper} compound {La$_4$Ni$_3$O$_{10}$}}.
\bjtitle{$ACS$ $Appl.$ $Electron.$ $Mater.$}
\bvolume{3}(\bissue{6}),
\bfpage{2671}--\blpage{2684}
(\byear{2021})
\end{barticle}
\endbibitem

%%% 40
\bibitem[\protect\citeauthoryear{Zhang et~al.}{2025}]{327damage}
\begin{botherref}
\oauthor{\bsnm{Zhang}, \binits{Y.}}, et al.:
Damage of bilayer structure in La$_3$Ni$_2$O$_{7-\delta}$ induced by high
  pO$_2$ annealing
(2025).
\url{https://arxiv.org/abs/2502.01501}
\end{botherref}
\endbibitem

%%% 41
\bibitem[\protect\citeauthoryear{Shi et~al.}{2025}]{La327.CXH}
\begin{barticle}
\bauthor{\bsnm{Shi}, \binits{M.}}, \betal:
\batitle{Prerequisite of superconductivity: {SDW} rather than tetragonal
  structure in double-layer {La$_3$Ni$_2$O$_{7-x}$}}.
\bjtitle{$Nat.$ $Commun.$}
\bvolume{16},
\bfpage{9141}
(\byear{2025})
\end{barticle}
\endbibitem

%%% 42
\bibitem[\protect\citeauthoryear{Liu et~al.}{2025}]{LIU2025125528}
\begin{barticle}
\bauthor{\bsnm{Liu}, \binits{Q.}}, \betal:
\batitle{Emergence of tetragonal phase in oxygen-annealed {Co}-doped
  {La$_3$Ni$_2$O$_{7+\delta}$}}.
\bjtitle{$J.$ $Solid$ $State$ $Chem.$}
\bvolume{351},
\bfpage{125528}
(\byear{2025})
\end{barticle}
\endbibitem

%%% 43
\bibitem[\protect\citeauthoryear{Ran et~al.}{2024}]{2024AMI.Cava}
\begin{barticle}
\bauthor{\bsnm{Ran}, \binits{G.}}, \betal:
\batitle{Is {La$_3$Ni$_{2}$O$_{6.5}$} a bulk superconducting nickelate?}
\bjtitle{$ACS$ $Appl.$ $Mater.$ $Interfaces$}
\bvolume{16}(\bissue{49}),
\bfpage{66857}--\blpage{66864}
(\byear{2024})
\end{barticle}
\endbibitem

%%% 44
\bibitem[\protect\citeauthoryear{Chen et~al.}{2024}]{1313.jacs}
\begin{barticle}
\bauthor{\bsnm{Chen}, \binits{X.}}, \betal:
\batitle{Polymorphism in the {Ruddlesden--Popper} nickelate
  {La$_3$Ni$_2$O$_7$}: discovery of a hidden phase with distinctive layer stacking}.
\bjtitle{$J.$ $Am.$ $Chem.$ $Soc.$}
\bvolume{146}(\bissue{6}),
\bfpage{3640}--\blpage{3645}
(\byear{2024})
\end{barticle}
\endbibitem

%%% 45
\bibitem[\protect\citeauthoryear{Wang et~al.}{2024}]{1313.XWW}
\begin{barticle}
\bauthor{\bsnm{Wang}, \binits{H.}},
\bauthor{\bsnm{Chen}, \binits{L.}},
\bauthor{\bsnm{Rutherford}, \binits{A.}},
\bauthor{\bsnm{Zhou}, \binits{H.}},
\bauthor{\bsnm{Xie}, \binits{W.}}:
\batitle{Long-range structural order in a hidden phase of {Ruddlesden--Popper}
  bilayer nickelate {La$_3$Ni$_2$O$_7$}}.
\bjtitle{$Inorg.$ $Chem.$}
\bvolume{63}(\bissue{11}),
\bfpage{5020}--\blpage{5026}
(\byear{2024})
\end{barticle}
\endbibitem

%%% 46
\bibitem[\protect\citeauthoryear{Puphal et~al.}{2024}]{1313.prl}
\begin{barticle}
\bauthor{\bsnm{Puphal}, \binits{P.}}, \betal:
\batitle{Unconventional crystal structure of the high-pressure superconductor
  {${\mathrm{La}}_{3}{\mathrm{Ni}}_{2}{\mathrm{O}}_{7}$}}.
\bjtitle{$Phys.$ $Rev.$ $Lett.$}
\bvolume{133},
\bfpage{146002}
(\byear{2024})
\end{barticle}
\endbibitem

%%% 47
\bibitem[\protect\citeauthoryear{Jiao et~al.}{2024}]{Jiao2024.PC}
\begin{barticle}
\bauthor{\bsnm{Jiao}, \binits{K.}}, \betal:
\batitle{Enhanced conductivity in {Sr} doped {La$_3$Ni$_2$O$_{7-\delta}$} with
  high-pressure oxygen annealing}.
\bjtitle{$Physica$ $C$}
\bvolume{621},
\bfpage{1354504}
(\byear{2024})
\end{barticle}
\endbibitem

%%% 48
\bibitem[\protect\citeauthoryear{Feng et~al.}{2024}]{Feng2024.PRB}
\begin{barticle}
\bauthor{\bsnm{Feng}, \binits{J.}}, \betal:
\batitle{Unaltered density wave transition and pressure-induced signature of
  superconductivity in {Nd-doped
  ${\mathrm{La}}_{3}{\mathrm{Ni}}_{2}{\mathrm{O}}_{7}$}}.
\bjtitle{$Phys.$ $Rev.$ $B$}
\bvolume{110},
\bfpage{100507}
(\byear{2024})
\end{barticle}
\endbibitem

%%% 49
\bibitem[\protect\citeauthoryear{Wang et~al.}{2025}]{wang2025.npj}
\begin{barticle}
\bauthor{\bsnm{Wang}, \binits{G.}}, \betal:
\batitle{Chemical versus physical pressure effects on the structure transition
  of bilayer nickelates}.
\bjtitle{$npj$ $Quantum$ $Mater.$}
\bvolume{10}(\bissue{1}),
\bfpage{1}
(\byear{2025})
\end{barticle}
\endbibitem

%%% 50
\bibitem[\protect\citeauthoryear{Rodriguez-Carvajal et~al.}{1991}]{JPCM1991}
\begin{barticle}
\bauthor{\bsnm{Rodriguez-Carvajal}, \binits{J.}},
\bauthor{\bsnm{Fernandez-Diaz}, \binits{M.}},
\bauthor{\bsnm{Martinez}, \binits{J.}}:
\batitle{Neutron diffraction study on structural and magnetic properties of
  {${\mathrm{La}}_{2}$${\mathrm{Ni}}$${\mathrm{O}}_{4}$}}.
\bjtitle{$J.$ $Phys.$ $Condens.$ $Matter$}
\bvolume{3}(\bissue{19}),
\bfpage{3215}
(\byear{1991})
\end{barticle}
\endbibitem

%%% 51
\bibitem[\protect\citeauthoryear{Aguadero et~al.}{2006}]{B605886H}
\begin{barticle}
\bauthor{\bsnm{Aguadero}, \binits{A.}}, \betal:
\batitle{In situ high temperature neutron powder diffraction study of
  oxygen-rich {La$_2$NiO$_{4+\delta}$} in air: correlation with the electrical
  behaviour}.
\bjtitle{$J.$ $Mater.$ $Chem.$}
\bvolume{16},
\bfpage{3402}--\blpage{3408}
(\byear{2006})
\end{barticle}
\endbibitem

%%% 52
\bibitem[\protect\citeauthoryear{Brown and Altermatt}{1985}]{BVS}
\begin{barticle}
\bauthor{\bsnm{Brown}, \binits{I.D.}},
\bauthor{\bsnm{Altermatt}, \binits{D.}}:
\batitle{{Bond-valence parameters obtained from a systematic analysis of the
  Inorganic Crystal Structure Database}}.
\bjtitle{$Acta.$ $Crystollogr.$ $B$}
\bvolume{41}(\bissue{4}),
\bfpage{244}--\blpage{247}
(\byear{1985})
\end{barticle}
\endbibitem

%%% 53
\bibitem[\protect\citeauthoryear{Shklovskii and
  Efros}{1984}]{shklovskii2013electronic}
\begin{bbook}
\bauthor{\bsnm{Shklovskii}, \binits{B.I.}},
\bauthor{\bsnm{Efros}, \binits{A.L.}}:
\bbtitle{Electronic Properties of Doped Semiconductors}
vol. \bseriesno{45}.
\bpublisher{Springer},
\blocation{Berlin}
(\byear{1984})
\end{bbook}
\endbibitem

%%% 54
\bibitem[\protect\citeauthoryear{Liu et~al.}{2023}]{Liu2023.SCP.WangM}
\begin{barticle}
\bauthor{\bsnm{Liu}, \binits{Z.}}, \betal:
\batitle{Evidence for charge and spin density waves in single crystals of
  {La$_3$Ni$_2$O$_7$} and {La$_3$Ni$_2$O$_6$}}.
\bjtitle{$Sci.$ $China$ $Phys.$ $Mech.$ $Astron.$}
\bvolume{66},
\bfpage{217411}
(\byear{2023})
\end{barticle}
\endbibitem

%%% 55
\bibitem[\protect\citeauthoryear{Mahajan
  et~al.}{1994}]{Mahajan1994.PRL.Zn-doping}
\begin{barticle}
\bauthor{\bsnm{Mahajan}, \binits{A.V.}},
\bauthor{\bsnm{Alloul}, \binits{H.}},
\bauthor{\bsnm{Collin}, \binits{G.}},
\bauthor{\bsnm{Marucco}, \binits{J.F.}}:
\batitle{$^{89}\mathrm{Y}$ {NMR} probe of {Zn} induced local moments in
  {YBa$_2$(Cu$_{1-y}$Zn$_y$)$_3$O$_{6+x}$}}.
\bjtitle{$Phys.$ $Rev.$ $Lett.$}
\bvolume{72},
\bfpage{3100}--\blpage{3103}
(\byear{1994})
\end{barticle}
\endbibitem

%%% 56
\bibitem[\protect\citeauthoryear{Zagoulaev
  et~al.}{1995}]{1995PhysRevB.Zagolaev}
\begin{barticle}
\bauthor{\bsnm{Zagoulaev}, \binits{S.}},
\bauthor{\bsnm{Monod}, \binits{P.}},
\bauthor{\bsnm{J\'egoudez}, \binits{J.}}:
\batitle{Magnetic and transport properties of {${\mathrm{Zn}}$}-doped
  {${\mathrm{YBa}}_{2}$${\mathrm{Cu}}_{3}$${\mathrm{O}}_{7}$} in the normal
  state}.
\bjtitle{$Phys.$ $Rev.$ $B$}
\bvolume{52},
\bfpage{10474}--\blpage{10487}
(\byear{1995})
\end{barticle}
\endbibitem

%%% 57
\bibitem[\protect\citeauthoryear{Yilmaz et~al.}{2025}]{yilmaz2025}
\begin{botherref}
\oauthor{\bsnm{Yilmaz}, \binits{H.}}, et al.:
Semiconducting character in {Ruddlesden-Popper bilayer nickelate
  Y$_{0.5}$Sr$_{2.5}$Ni$_{2-x}$Al$_{x}$O$_7$}
(2025).
\url{https://arxiv.org/html/2503.07861}
\end{botherref}
\endbibitem

%%% 58
\bibitem[\protect\citeauthoryear{Ishida
  et~al.}{1996}]{Ishida1996.PRL.Al-doping}
\begin{barticle}
\bauthor{\bsnm{Ishida}, \binits{K.}},
\bauthor{\bsnm{Kitaoka}, \binits{Y.}},
\bauthor{\bsnm{Yamazoe}, \binits{K.}},
\bauthor{\bsnm{Asayama}, \binits{K.}},
\bauthor{\bsnm{Yamada}, \binits{Y.}}:
\batitle{Al {NMR} probe of local moments induced by an {Al} impurity in
  {High-${T}_{c}$} cuprate
  {${\mathrm{La}}_{1.85}{\mathrm{Sr}}_{0.15}{\mathrm{CuO}}_{4}$}}.
\bjtitle{$Phys.$ $Rev.$ $Lett.$}
\bvolume{76},
\bfpage{531}--\blpage{534}
(\byear{1996})
\end{barticle}
\endbibitem

%%% 59
\bibitem[\protect\citeauthoryear{Tarascon
  et~al.}{1988}]{tarascon1988structural}
\begin{barticle}
\bauthor{\bsnm{Tarascon}, \binits{J.}}, \betal:
\batitle{Structural and physical properties of the metal ({$M$}) substituted
  {${\mathrm{YBa}}_{2}$${\mathrm{Cu}}_{3-x}$$M_{x}$${\mathrm{O}}_{7}$}
  perovskite}.
\bjtitle{$Phys.$ $Rev.$ $B$}
\bvolume{37}(\bissue{13}),
\bfpage{7458}
(\byear{1988})
\end{barticle}
\endbibitem

%%% 60
\bibitem[\protect\citeauthoryear{Anderson}{1959}]{anderson1959theory}
\begin{barticle}
\bauthor{\bsnm{Anderson}, \binits{P.W.}}:
\batitle{Theory of dirty superconductors}.
\bjtitle{$J.$ $Phys.$ $Chem.$ $Solids$}
\bvolume{11}(\bissue{1-2}),
\bfpage{26}--\blpage{30}
(\byear{1959})
\end{barticle}
\endbibitem

%%% 61
\bibitem[\protect\citeauthoryear{Millis et~al.}{1988}]{millis1988inelastic}
\begin{barticle}
\bauthor{\bsnm{Millis}, \binits{A.}},
\bauthor{\bsnm{Sachdev}, \binits{S.}},
\bauthor{\bsnm{Varma}, \binits{C.}}:
\batitle{Inelastic scattering and pair breaking in anisotropic and isotropic
  superconductors}.
\bjtitle{$Phys.$ $Rev.$ $B$}
\bvolume{37}(\bissue{10}),
\bfpage{4975}
(\byear{1988})
\end{barticle}
\endbibitem

%%% 62
\bibitem[\protect\citeauthoryear{Toby and Von~Dreele}{2013}]{toby2013gsas}
\begin{barticle}
\bauthor{\bsnm{Toby}, \binits{B.H.}},
\bauthor{\bsnm{Von~Dreele}, \binits{R.B.}}:
\batitle{{GSAS-II}: the genesis of a modern open-source all purpose
  crystallography software package}.
\bjtitle{$J.$ $Appl.$ $Crystallogr.$}
\bvolume{46}(\bissue{2}),
\bfpage{544}--\blpage{549}
(\byear{2013})
\end{barticle}
\endbibitem

%%% 63
\bibitem[\protect\citeauthoryear{Mao et~al.}{1986}]{mao1986calibration}
\begin{barticle}
\bauthor{\bsnm{Mao}, \binits{H.}},
\bauthor{\bsnm{Xu}, \binits{J.}},
\bauthor{\bsnm{Bell}, \binits{P.}}:
\batitle{Calibration of the ruby pressure gauge to 800 kbar under
  quasi-hydrostatic conditions}.
\bjtitle{$J.$ $Geophys.$ $Res.$ $Solid$ $Earth$}
\bvolume{91}(\bissue{B5}),
\bfpage{4673}--\blpage{4676}
(\byear{1986})
\end{barticle}
\endbibitem

\end{thebibliography}

\end{document}